\documentclass[twocolumn,prapplied,floatfix]{revtex4}
\usepackage{amsmath}
\usepackage{amssymb}
\usepackage{amsmath}
\usepackage{graphicx}
\usepackage{tikz}
\usetikzlibrary{quantikz}

\begin{document}

\title{Rescaling interactions for quantum control}
\author{Gaurav Bhole}
\author{Takahiro Tsunoda}
\author{Peter J. Leek}
\author{Jonathan A. Jones}
\affiliation{Clarendon Laboratory, University of Oxford, Parks Road, Oxford OX1 3PU, United Kingdom}
\date{\today}

\begin{abstract}
A powerful control method in experimental quantum computing is the use of spin echoes, employed to select a desired term in the system's internal Hamiltonian, while refocusing others.  Here we address a more general problem, describing a method to not only turn on and off particular interactions but also to rescale their strengths so that we can generate any desired effective internal Hamiltonian. We propose an algorithm based on linear programming for achieving time-optimal rescaling solutions in fully coupled systems of tens of qubits, which can be modified to obtain near time-optimal solutions for rescaling systems with hundreds of qubits.
\end{abstract}

\maketitle

%\section{Introduction}
In the circuit model of quantum computation quantum algorithms are implemented by applying a network of quantum logic gates to a set of qubits. Typically these qubits are defined by the internal Hamiltonian of the underlying system, while logic gates are applied using external control fields. In some cases, such as trapped ion implementations \cite{Brown2016}, the free evolution under the internal Hamiltonian corresponds to single-qubit $z$-rotations, and two-qubit gates are generated as and when required, for example by coupling internal and external degrees of freedom \cite{Schaefer2018}. In other cases, exemplified by nuclear magnetic resonance (NMR), the internal Hamiltonian contains two-qubit $zz$-interactions, and two-qubit gates are generated by free evolution \cite{Jones2011}. Many other proposed implementations \cite{Barenco1995,Kane1998,Loss1998,Brennen1999,Jaksch1999,Knill2001,Leuenberger2001,Harneit2002,Suter2002,Chen2019} fit into one of these two paradigms or a mixture \cite{Parra-Rodriguez2018}: for example NV centres are NMR like when controlling the spins around a single centre \cite{Wu2019}, but generate long-range two-qubit gates on demand \cite{Moehring2007}. Superconducting qubits commonly use gates generated on demand \cite{Arute2019}, but can also be tuned to generate always-on two-qubit interactions \cite{Gambetta2017}.

With always-on two-qubit interactions free evolution provides a universal quantum logic gate when combined with single-qubit rotations, but this gate is not a convenient one as it corresponds to a complex pattern of evolutions. For this reason methods such as spin echoes \cite{Hahn1950} are used to replace the background Hamiltonian with a more convenient average Hamiltonian, in which desired interactions are isolated while undesirable interactions are refocused. As a concrete example system we consider NMR spin systems, but equivalent Hamiltonians can be found in other quantum information processing (QIP) implementations.

Previous studies in NMR \cite{FreemanSCbook,Jones1999,Leung2000,Leung2002} have largely concentrated on methods for refocusing all the interactions, or for isolating one single interaction while refocusing everything else. However, a more general problem is to \textit{rescale} the size of interactions in the Hamiltonian, to produce a desired effective Hamiltonian. Here we describe a method for finding a rescaling sequence with the shortest possible total time, and with a fairly small number of echo pulses.  In its simplest form this is practical for systems of up to about 20 qubits, but for larger systems we have developed a pragmatic method for finding short, although not perfectly optimal, rescaling sequences, which works with more than 100 qubits, meaning that the method could be of use for practically useful quantum computation in scalable platforms. Related ideas have been explored in systems with other similar Hamiltonians \cite{Parra-Rodriguez2018,Hayes2014,Welch2014}.

The initial sections below introduce the terminology of spin echoes and reprise previous results for refocusing sequences. This is followed by a description of our new rescaling method with some sample applications in systems with a small number of qubits. Finally we show how random sampling can be used to extend these methods to much larger systems.

\section{Spin echoes}
The background Hamiltonian of an NMR spin system consists of one spin interactions (resonance offsets) and two spin interactions (J-couplings). Consider a system of $q$ spins where the $i^\text{th}$ spin has a resonance offset $\Omega_i$ and the pair of spins $i$ and $j$ have a J-coupling frequency $\omega_{ij}$. The Hamiltonian for this system in the weak coupling limit is,
\begin{equation}
\mathcal{H} = \sum_{i} \Omega_i I_{z}^i + \sum_{i<j} \omega_{ij} I_{z}^iI_{z}^j,
\end{equation}
where, following NMR notation, factors of $\hbar$ have been dropped, and $I_z^i = \sigma_z^i /2$ is the Pauli spin-$1/2$ operator acting on the $i^{\text{th}}$ spin. In practice some of these interactions could either be zero or set to zero, allowing them to be dropped. For example, working in a suitable rotating frame allows some of the resonance offsets to be set to zero, while many J-couplings can be negligible in partially coupled spin systems. However, for generality and completeness, we shall initially consider fully coupled systems with $q$ non-zero resonance-offsets and $p = q(q-1)/2$ non-zero J-couplings between the $p$ pairs of spins.

During a period $\tau$ of free evolution each spin evolves under all the $q+p$ interactions, given by the propagator $U = \exp{(-\mathrm{i} \mathcal{H}\tau)}$. Since this Hamiltonian is diagonal in the chosen $z$-basis, all terms in the Hamiltonian commute, and the one-spin and two-spin evolutions can be summarised by the acquired phases
\begin{equation}
\Phi_i=\Omega_i\tau,\quad\phi_{ij}=\omega_{ij}\tau.
\end{equation}
To sculpt the effective Hamiltonian into the desired form it is necessary to isolate the spin interactions which we want while suppressing the unwanted interactions. This essentially requires controlling the evolution of the spins such that the unwanted interactions finally acquire a phase of zero while letting the required interactions evolve to reach the desired values. %Although this manipulation can be achieved by a variety of techniques, the most simple and elegant approach relies on spin echoes.

A spin echo is a period of free evolution interrupted by $\pi$ rotations in the middle of the evolution period. The notation $\pi^i$ is used to denote a pulse which causes a $180$ degree rotation on spin $i$, about the $x$-axis unless otherwise stated. The effect of a pair of $\pi^i$ rotations is to negate the effective frequency $\Omega_i$ of the spin for the time period between the two pulses. Thus, the sequence $\tau\,\pi^i\,\tau\,\pi^i$, where time periods indicate free evolution under the internal Hamiltonian, will refocus the offset $\Omega_i$ as the phase $\Phi_i$ acquired during the first period of evolution gets nullified by the phase $-\Phi_i$ acquired in the second half.

Clearly, the one-spin interaction of any given spin $i$ is only affected by $\pi$ rotations applied to spin $i$, and so individual interactions can be controlled independently. However, for two-spin interactions, a $\pi$ rotation applied to either spin $i$ or $j$ reverses the frequency $\omega_{ij}$ while a simultaneous $\pi$ rotation applied on both spins $i$ and $j$ leaves $\omega_{ij}$ unchanged. A sequence $\tau\,\pi^i\,\tau\,\pi^i$ will thus refocus $\Omega_i$ and $\omega_{ij}$, whereas a sequence $\tau\,\pi^{i,j}\,\tau\,\pi^{i,j}$ will refocus both $\Omega_i$ and $\Omega_j$ but not $\omega_{ij}$. In this manner, the two-spin interactions can be controlled, but this control cannot be achieved independently from one-spin interactions.

A general spin echo sequence comprises a series of free evolution time periods $\tau_m$, sometimes called delays, separated by $\pi$ pulses applied to one or more spins. As long as the total number of $\pi$ pulses applied to a given spin is even the overall evolution can still be summarised by a set of phases, but now
\begin{equation}
\Phi_i=\Omega_i\sum_m S^i_m\tau_m,\quad\phi_{ij}=\omega_{ij}\sum_mS^i_mS^j_m\tau_m,\label{eq:phi}
\end{equation}
where $S$ is a sign matrix, containing only the elements $\pm1$, with a sign change whenever a $\pi$ pulse is applied to spin $i$. For convenience we will also refer to the two-qubit sign matrix $S^{ij}_{m}=S^i_m S^j_m$, although this is obviously not independent from the one-qubit matrix. The complete sign matrix can be obtained by combining the one- and two-spin matrices.

\section{Refocusing}
Methods for removing all the interactions (sometimes called decoupling), or for isolating one single interaction while refocusing everything else, have been widely studied. The most effective methods to achieve this rely on choosing sign matrices whose rows are taken from Walsh--Hadamard matrices \cite{Jones1999,Leung2000,Leung2002}, so that each row is a Walsh function \cite{Beauchamp1984}. These matrices differ from other Hadamard matrices in that they are only defined for dimensions equal to a power of 2, the rows are not normalised, and the ordering of the rows is different.

A Walsh function $W_n$ is defined by a vector with length equal to a power of 2 and with all the entries equal to $\pm1$.  For $W_0$ all the entries are $+1$, while for every other $W_n$ half the entries are $+1$ and half are $-1$, with the entries arranged such that there are $n$ regularly spaced sign changes along the vector. For example the 4 by 4 Walsh--Hadamard matrix contains the four rows
\begin{equation}
\begin{pmatrix}W_0\\W_1\\W_2\\W_3\end{pmatrix}=\begin{pmatrix}+1&+1&+1&+1\\+1&+1&-1&-1\\+1&-1&-1&+1\\+1&-1&+1&-1\end{pmatrix}
\end{equation}
Strictly the name of the Walsh function must specify the number of columns as well as the number of sign changes, but this is left implicit here: the number of columns is equal to the smallest power of 2 larger than the highest Walsh number considered.

In a system of three spins it is possible to remove all three one-spin and all three two-spin interactions by using four equal time periods $\tau$ and a sign matrix obtained by choosing $S^i=W_i$, avoiding $W_0$. This relies on two key properties of Walsh functions. Firstly all Walsh functions except $W_0$ contain an equal number of $\pm1$ values, and so all one-qubit interactions will be refocused when equal length time periods are used. Secondly the product of two Walsh functions is itself a Walsh function \cite{Beauchamp1984}, defined by
\begin{equation}
W_p\circ W_q=W_{p\oplus q}
\end{equation}
where $\circ$ indicates element wise multiplication, sometimes called the Schur product \cite{Lynn1964}, and $\oplus$ indicates bitwise addition modulo two. Thus all two-qubit interactions will also be refocused.

A decoupling network is easily modified \cite{Jones1999,Leung2000} to retain a single interaction: to retain a one-spin interaction $\Omega_i$ use $S_i=W_0$ for this spin, while to retain a coupling $\omega_{ij}$ set $S_i=S_j$ so that $S_{ij}=W_0$.  To take a concrete example the coupling $\omega_{12}$ can be isolated in a three spin system by choosing $S_1=S_2=W_1$ and $S_3=W_2$. The $\pi$ pulses required can be deduced by applying a pulse to a spin whenever the corresponding row of $S$ changes sign, including a final $\pi$ pulse if the $S$ row ends in $-1$, giving the sequence
\begin{equation}
\tau\, \pi^3\, \tau\, \pi^{1,2}\, \tau\, \pi^3\, \tau\, \pi^{1,2}.
\end{equation}
Note that $\omega_{12}$ evolves with sign $+1$ at every stage, and so is retained at full strength. The total evolution time required is given by $4\tau=\phi_{12}/\omega_{12}$. If this expression gives a negative time then this can be resolved by applying additional $\pi$ pulses to one spin at the beginning and end of the sequence to negate the evolution frequency.

Now consider how this approach scales to a system of $q$ spins.  Retaining a single interaction can be done efficiently: the number of time periods required is given by the smallest power of 2 larger than $q$, which is upper bounded by $2q$, while the number of individual $\pi$ pulses required is clearly less than $2q^2$, which corresponds to applying a pulse to every spin after every time period.  A more careful analysis (see Appendix~\ref{app:pulsecount}) shows that only around $q^2/2$ pulses are required, which is still $O(q^2)$ but with a smaller pre-factor. As the single interaction is retained at full strength this is also a minimum time solution.

\section{Rescaling}\label{sec:Rescaling}
So far, we have only considered retaining a single one-spin or two-spin interaction while refocusing the remainder. However, a more general problem is to \textit{rescale} the size of interactions in the Hamiltonian. In other words, we desire to achieve a certain set of non-zero target phases for \textit{all} the spin interactions. The obvious approach is just to place spin echo sequences which isolate the individual interactions back to back. As there are a total of $r=q+p=q(q+1)/2$ single-spin and two-spin interactions to be considered it is clear that the number of time periods is $O(q^3)$, and the number of pulses is $O(q^4)$.  The total time required is given by the sum of the times required to evolve under each individual interaction,
\begin{equation}
T=\sum_i\left|\frac{\Phi_i}{\Omega_i}\right|+\sum_{i<j}\left|\frac{\phi_{ij}}{\omega_{ij}}\right|.\label{eq:Tn}
\end{equation}
This na\"ive approach is expensive, both in time and the number of $\pi$ pulses. We thus want to find a more efficient sequence, by carrying out as many evolutions in parallel as far as possible. Although this might sound challenging, we propose here a straightforward way to achieve this using linear programming.  This new approach also greatly reduces the number of pulses and time periods required.

%\subsection{Rescaling in smaller spin systems ($q<12$)}
\subsection{Setting up the problem}
We consider a system of $q$ coupled spins described by $r = q(q+1)/2$ one- and two-spin interactions. Our aim is to rescale all $r$ interactions simultaneously such that they reach the desired target phases, which for generality we assume to be all different. We begin by constructing an overcomplete Walsh basis by building a one-spin sign matrix with the rows given by Walsh functions numbered $2^j$, where $j = 0,1,\dots (q-1)$.  Next, we use this matrix to construct the two-spin sign matrix by taking products of corresponding rows in the one-spin matrix.

Combining these by stacking the two matrices together gives the complete sign matrix $S$ of $r$ rows and $s=2^q$ columns. The single-spin functions correspond to binary numbers with precisely one bit set, while the two-spin functions correspond to binary numbers with precisely two bits set. As these numbers are all distinct it is guaranteed that the complete sign matrix has enough flexibility to permit every interaction to be controlled separately. This is quite different from refocusing sequences, where many functions are repeated.

\subsection{Linear Programming}
This overcomplete Walsh basis guarantees that solutions to Eqn.~$\ref{eq:phi}$ can be found for any target values of $\{\Phi_i\}$ and $\{\phi_{ij}\}$ by choosing $2^q$ appropriate values of $\{\tau_m\}$, but it is not obvious how these can be found. As the basis is overcomplete, multiple solutions will exist but these can be distinguished by requiring that all the times $\{\tau_m\}$ must be non-negative and by preferring the solutions with the shortest value of total time $T = \sum_{m}\tau_m$. These criteria for desirable solutions suggest a powerful method, namely linear programming \cite{Bland1981}.

The general linear programming problem varies some inputs (here the times $\{\tau_m\}$) seeking to minimize some linear function of these inputs (here the total time $T$) subject to a number of equality constraints (here Eqn. $\ref{eq:phi}$) and inequality constraints (here, that each $\tau_m \ge 0$). We adopted a simple approach, using the inbuilt Matlab function \texttt{linprog}.

It is important to consider the computational complexity of linear programming, as this determines how the time required to find a solution scales with the number of qubits $q$. The precise computational complexity of linear programming is known to be poorly defined, depending on both the algorithm used and the details of the problem, and with the worst case behaviour being very different from the typical case \cite{Bland1981}. The Matlab function \texttt{linprog} has a choice of two algorithms: the original simplex algorithm developed by Dantzig \cite{Dantzig1982}, and a more modern interior point algorithm \cite{Adler1989}, which are briefly discussed in Appendix~\ref{app:lp}. Both algorithms typically have computational complexity between $O(n^2)$ and $O(n^3)$, where $n$, the dimension of the problem, can be taken as the sum of the number of rows and columns in the constraint matrix, so that here $n=r+s\approx2^q$.  We investigated this question experimentally by simply timing the program. %using each of the two possible algorithms.
All results are for the simplex algorithm unless otherwise stated.

\subsection{Extracting solutions}
As the linear programming algorithm is fundamentally trying to minimise $T$, subject to the positivity constraint and the target phases, the algorithm prefers solutions where many of the $\{\tau_m\}$ are zero. (This is not specific to this problem, but is a general feature of linear programming solutions \cite{NumRec1992}.) The linear programming solution has at most only as many non-zero times as the number of constraints $r$ in the problem. If the problem involves extensive refocusing rather than rescaling then solutions with an even smaller number of non-zero times can be found.

It is obviously not necessary to explicitly implement the time periods of length $0$, and so the overcomplete sign matrix, $S$, can be replaced by a reduced matrix, $R$, by selecting only $r$ or fewer columns from $S$ which correspond to non-zero evolution times.

\subsection{Optimizing the solutions}
One subtlety is that the order of columns in the $R$ matrix does not affect the phases produced, but different orderings of these columns can lead to pulse sequences with different numbers of pulses. As minimising the number of pulses is desirable it is useful to explore different permutations of the $R$ matrix, seeking for the arrangement which gives the smallest number of sign changes.

If the matrix is not too large then exhaustive permutation can be practical, but in larger cases it is more sensible to select a number of random permutations and keep the best one found. Experience so far suggests that different permutations can require numbers of pulses that differ by a factor of around two. We also find that the pulse pattern corresponding to the original $R$ matrix is typically relatively good, although rarely the absolute best.

\subsection{Building the pulse sequence}
From this optimal reduced matrix $R$, a pattern of pulses can be generated by applying a $\pi$ pulse to every spin whose sign changes. It is important to remember that all interactions start at $+1$ and must end at $+1$, which can be modelled by adding initial and final columns to $R$ containing entirely $+1$. These additional columns have evolution times set to zero, and so are not actually implemented, but the resulting sign changes make it necessary to apply pulses to some spins at the start and end of the sequence. This also ensures that the number of $\pi$ pulses applied to each spin is even, which is required to create true spin echoes.

As the reduced matrix has $r\approx q^2/2$ times the final pulse sequence will have $O(q^2)$ time periods and $O(q^3)$ individual $\pi$ pulses, which is a very significant improvement on na\"ive methods.

\section{Example calculations}
We illustrate our method of rescaling with the help of two specific examples before returning to the general case, when we will consider the computational time required to find these optimal solutions.
\subsection{Homonuclear 3-spin fully coupled system}
First consider the example of iodotrifluoroethene \cite{Du2007}, $\rm C_2F_3I$, with the three $^{19}$F nuclei forming our spin system, in a magnetic field such that the Larmor frequency of $^{1}$H nuclei is 600\,MHz, with the excitation frequency set in the middle of the spectral range.
\begin{figure}[tb]
	\centering
	\includegraphics[width=65mm]{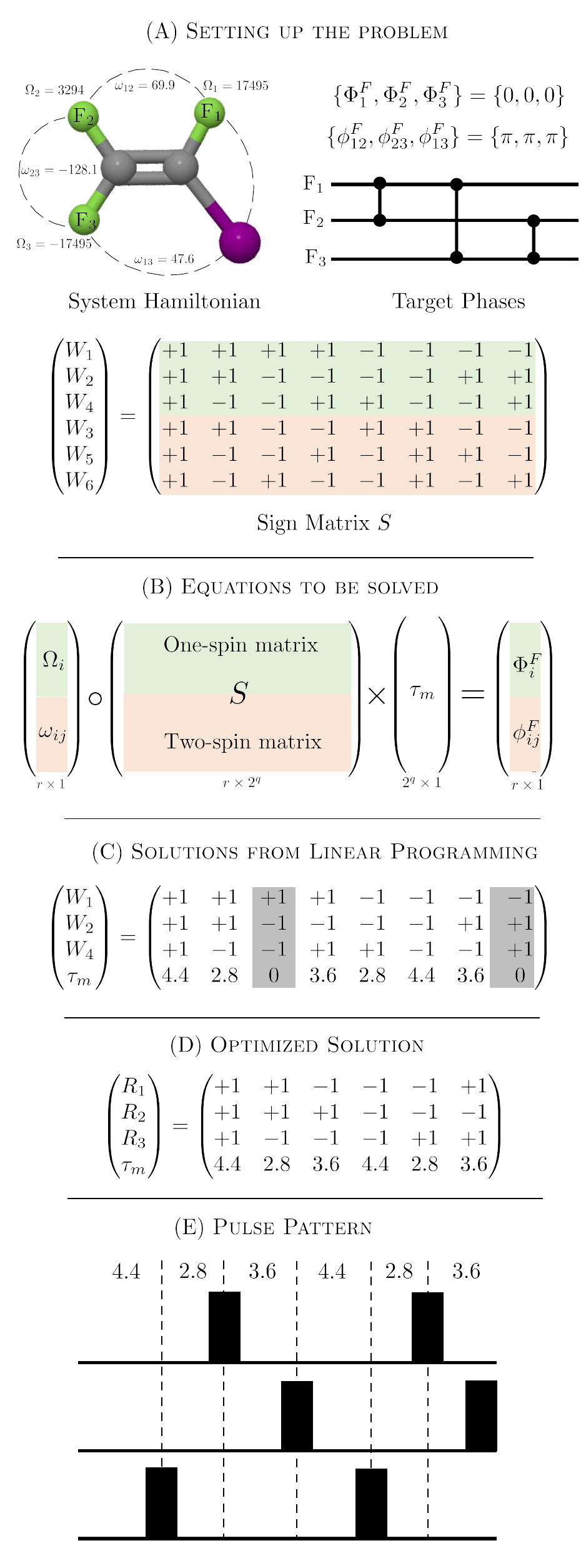}
	\caption{Implementation of our algorithm using the three $^{19}$F nuclei in iodotrifluoroethene, forming a three-spin full coupled homonuclear NMR system. All interactions are written in Hz and times in ms. The stages of the implementation are described in the main text.}
	\label{algo}
\end{figure}
The molecular structure and the Hamiltonian of the system is shown in Fig.~\ref{algo}\,(A).

This system has $r=6$ interactions altogether. Our aim here is to rescale all three of the two-spin interactions such that they have equal effective strength, while refocusing the three one-spin interactions. This is equivalent to achieving the target phases $\{\Phi_1, \Phi_2, \Phi_3\} = \{0, 0, 0\}$ and $\{\phi_{12}, \phi_{23}, \phi_{13}\} = \{\pi, \pi, \pi\}$.
The quantum circuit shown in Fig.~\ref{algo}\,(A) uses the coupling gate \cite{Knill2000,Bowdrey2005}, which is equivalent to a controlled-$Z$ gate, with the notation indicating the complete symmetry of a controlled gate implemented through $zz$ couplings, with no distinction between control and target spins.
Actual controlled-$Z$ gates would require additional single-spin $Z$ rotations, so that $\{\Phi_1, \Phi_2, \Phi_3\} = \{\pi, \pi, \pi\}$. These can be either be implemented directly in the pulse sequence \cite{Freeman1981,Morton2006a,Jones2013}, or tracked and corrected at the end \cite{Knill2000,Bowdrey2005}. Both approaches have been used in the literature, with tracking being particularly popular for fixed phases which are integer multiples of $\pi/2$, as seen here, as this is relatively simple in many experimental implementations.

Here we will consider both cases, beginning with tracking, so that the single-qubit target phases are all 0.  Note that even though the target phases are all 0 or $\pi$, the corresponding evolution times will be different as the one- and two-spin interactions have different values. Thus, we expect to require up to six time periods, consistent with $r=6$ interactions.

We begin by constructing an overcomplete basis from a one-spin sign matrix with three rows and eight columns, corresponding to the Walsh functions $\{W_1, W_2, W_4\}$, with the explicit form shown in Fig.~\ref{algo}\,(A).  The two-spin interaction matrix can be constructed by taking the three unique products of rows in the one-spin interaction matrix. The resulting set of equations is shown schematically in Fig.~\ref{algo}\,(B).  Here $\circ$ indicates an element wise (Schur) product as before, while $\times$  indicates a conventional matrix product.

As the sign matrix has more columns that rows there is no unique solution, but linear programming can be used to find a solution which minimises the total time while keeping all time periods positive. The program finds an optimal solution which uses only six columns as expected, by assigning zero time periods to two of the columns, as shown in Fig.~\ref{algo}\,(C).
The total time $T$ of the evolution is 21.6\,ms, which in this case is identical to the time required for a na\"ive implementation with sequential gates. This is a general rule for rescaling multiple couplings in three-spin systems, as it is impossible to control any coupling without affecting all the others.

However, even in this case, the resulting pulse sequence, requiring eight pulses, is much simpler than na\"ive sequential designs, which require 18 pulses. We can further improve this by permuting the columns of the reduced sign matrix $R$, as shown in Fig.~\ref{algo}\,(D), to find a solution with a smaller number of sign changes. From the reduced sign matrix, we can now calculate the pattern of $\pi$ pulses, Fig.~\ref{algo}\,(E), with a $\pi$ pulse applied wherever we encounter a sign change. Since there is one sign change in row $R_2$ it might seem that we need only one $\pi$ pulse. However, since the row ends in $-1$ we must apply another $\pi$ pulse at the end, leading to a total of six $\pi$ pulses.  Similarly, reduced sign matrix rows might in some cases begin with $-1$, and in this case it is necessary to apply a $\pi$ pulse before the first time period.

Next we consider the case where tracking is not used, so that the single-qubit target phases of $\pi$ need to be implemented directly.  In this case a na\"ive sequential approach would require a small additional evolution time divided up by 12 additional pulses. By contrast our efficient method can achieve this with no additional evolution time or pulses by slightly unbalancing the identical pairs of times seen in the case above.  It is always possible to do this as long as all the single-qubit offset frequencies are large in comparison with the two-qubit coupling strengths, which is usually the case and can always be achieved by changing the central excitation frequency.

We have assumed here that the $\pi$ pulses are implemented perfectly, that is there is no evolution under the drift Hamiltonian while the $\pi$ pulse is applied. This will be approximately true in a heteronuclear spin system, while in a homonuclear system equivalent pulses can be designed using techniques such as Gradient Ascent Pulse Engineering (GRAPE) \cite{Khaneja2005}.

\subsection{Homonuclear 4-spin linear chain}
\begin{figure}[htb]
	\centering
	\includegraphics[width=85mm]{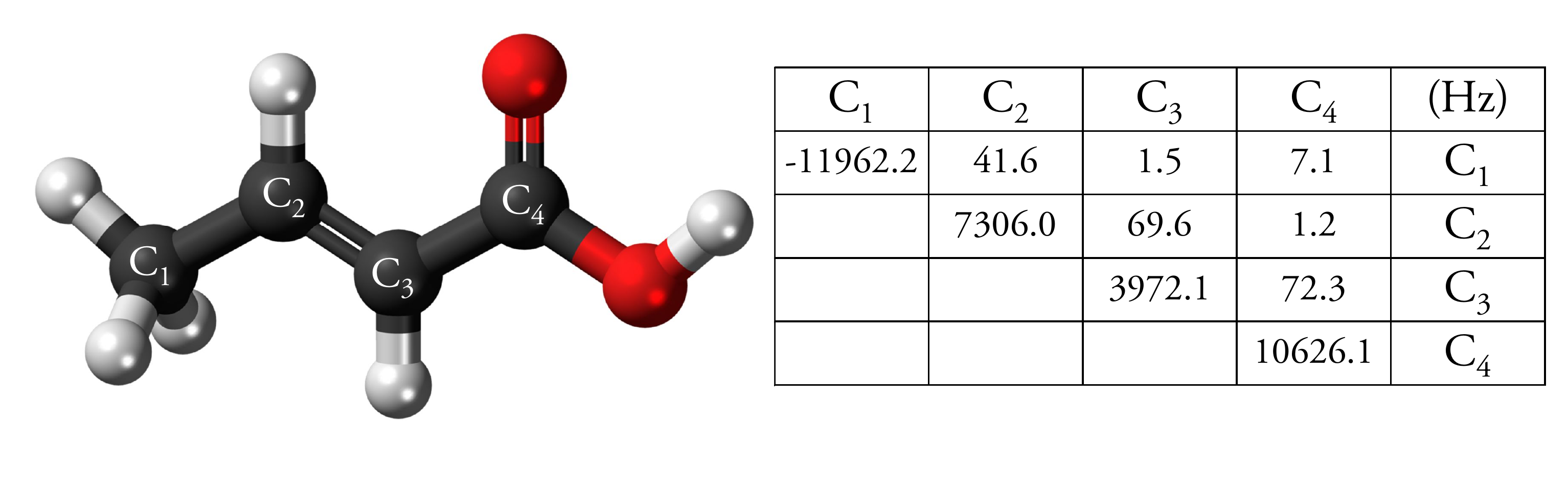}
	\caption{Molecular structure of fully labelled crotonic acid and the corresponding Hamiltonian for four $^{13}$C nuclei forming a four-spin chain like homonuclear NMR system.} %The diagonals in the Hamiltonian correspond to one-spin interactions or chemical shifts whereas the off diagonals represent the two-spin interactions or J-couplings.}
	\label{cccc}
\end{figure}

Next consider an NMR system provided by four $^{13}$C nuclei in labelled crotonic acid \cite{Boulant2002}. The four spins form a rough linear chain, with large one-spin interactions, moderate nearest neighbour couplings and significantly smaller long range couplings as shown in Fig \ref{cccc}.
Our typical target for this system is to refocus all the one-spin interactions and the three small couplings, but to retain the three nearest neighbour couplings. We also want to rescale the retained couplings such that they have the same target phases. We have a total of $10$ interactions from which we need to refocus $7$ while rescaling $3$. This can also be represented, up to single-spin phases, by the quantum circuit shown below.
\begin{equation}
	\begin{tikzcd}%[slice all]
		\qw &\ctrl{1} &\qw&\qw&\qw\\
		\qw &\control{} &\ctrl{1}&\qw &\qw\\
		\qw &\qw        &\control{}&\ctrl{1}&\qw\\
		\qw &\qw        &\qw&\control{}&\qw
	\end{tikzcd}
\end{equation}
In other words we want the target phases to be
\begin{equation}\{\Phi_1, \Phi_2, \Phi_3,\Phi_4\} = \{0, 0, 0,0\}\end{equation}
and
\begin{equation}\{\phi_{12}, \phi_{13}, \phi_{14}, \phi_{23}, \phi_{24}, \phi_{34}\} = \{\pi, 0, 0, \pi, 0, \pi\}.\end{equation}

We begin by constructing the overcomplete basis by building a sign matrix $S$ with 10 rows and 16 columns which correspond to the Walsh functions $\{W_1, W_2, W_4, W_8\}$ and their six distinct products, which are $\{W_3, W_5, W_6, W_9, W_{10}\}$. Running linear prediction and following the procedure as in the previous example, we now find a solution with 9 time periods
\begin{equation}
\begin{pmatrix}R_1\\R_2\\R_3\\R_4\\\tau_m\end{pmatrix}=\begin{pmatrix}+1&+1&-1&-1&-1&-1&+1&+1&+1\\+1&+1&+1&-1&-1&-1&-1&+1&+1\\+1&+1&+1&+1&-1&-1&-1&-1&-1\\+1&-1&-1&+1&+1&-1&-1&-1&+1\\3.5&1.3&1.8&3.0&1.8&3.0&1.8&1.7&1.3\end{pmatrix}
\label{R9}
\end{equation}
which can be implemented with 10 pulses. Note that since our target consists mostly of zeros the number of time periods required here is less than $r$. In this case the total time required is only 19.2\,ms, significantly shorter than the na\"ive sequential time of 26.1\,ms.

As a second example, we now consider the case where we only seek to implement the first and third controlled-$Z$ gates, so that the coupling phase are now
\begin{equation}\{\phi_{12}, \phi_{13}, \phi_{14}, \phi_{23}, \phi_{24}, \phi_{34}\} = \{\pi, 0, 0, 0, 0, \pi\}.\end{equation}
In this case only 6 time periods are required, although once again 10 pulses are needed to implement the desired sign changes. The total time required is now 12.0\,ms, which is identical to the time required to implement just the first gate by na\"ive methods.  Thus in this case our algorithm has found a solution which implements two gates in parallel, at no additional cost in total time. Clearly it is not possible to implement the gates any quicker than this, as the time required to implement the slowest gate on its own provides a firm lower bound. An upper bound is provided by the time required to implement each of the $r$ evolutions in sequence, Eq.~\ref{eq:Tn}, which is simply the sum of the times required for each gate.

\subsection{General solutions}
We have repeated calculations of this kind in a large variety of fictitious spin systems with increasing numbers of spins, up to $q=18$, and solutions have always been found. For larger values of $q$ these solutions are always more time-efficient than the sequential approach, and are usually far quicker. The greatest savings are found in cases where a moderate number of gates need to be implemented in parallel, and particularly when unused long-range couplings are significantly weaker than the couplings being controlled.

Until now we have described the problem as if there was a unique optimal solution which the linear programming locates. In fact there are multiple equivalent solutions, from among which the linear programming chooses one. These alternative solutions can be easily generated by permuting the columns of the $S$ matrix before running the algorithm, but as they all have the same number of individual time periods and take the same total time there is no good reason to do this.

We note in passing that these optimal solutions are only optimal for implementations containing only delays and $\pi$ pulses. If it is desired to implement an evolution corresponding to a weak long-range coupling then it may be quicker to use \textsc{swap} gates and related methods to implement long-range interactions through a chain of stronger short-range interactions \cite{Collins2000,Khaneja2007}. Even in such cases, however, the ideas described here can be used to assist in the design of such indirect gates.

\subsection{Computation time complexity}
\begin{figure}[tb]
	\centering
	\includegraphics[width=85mm]{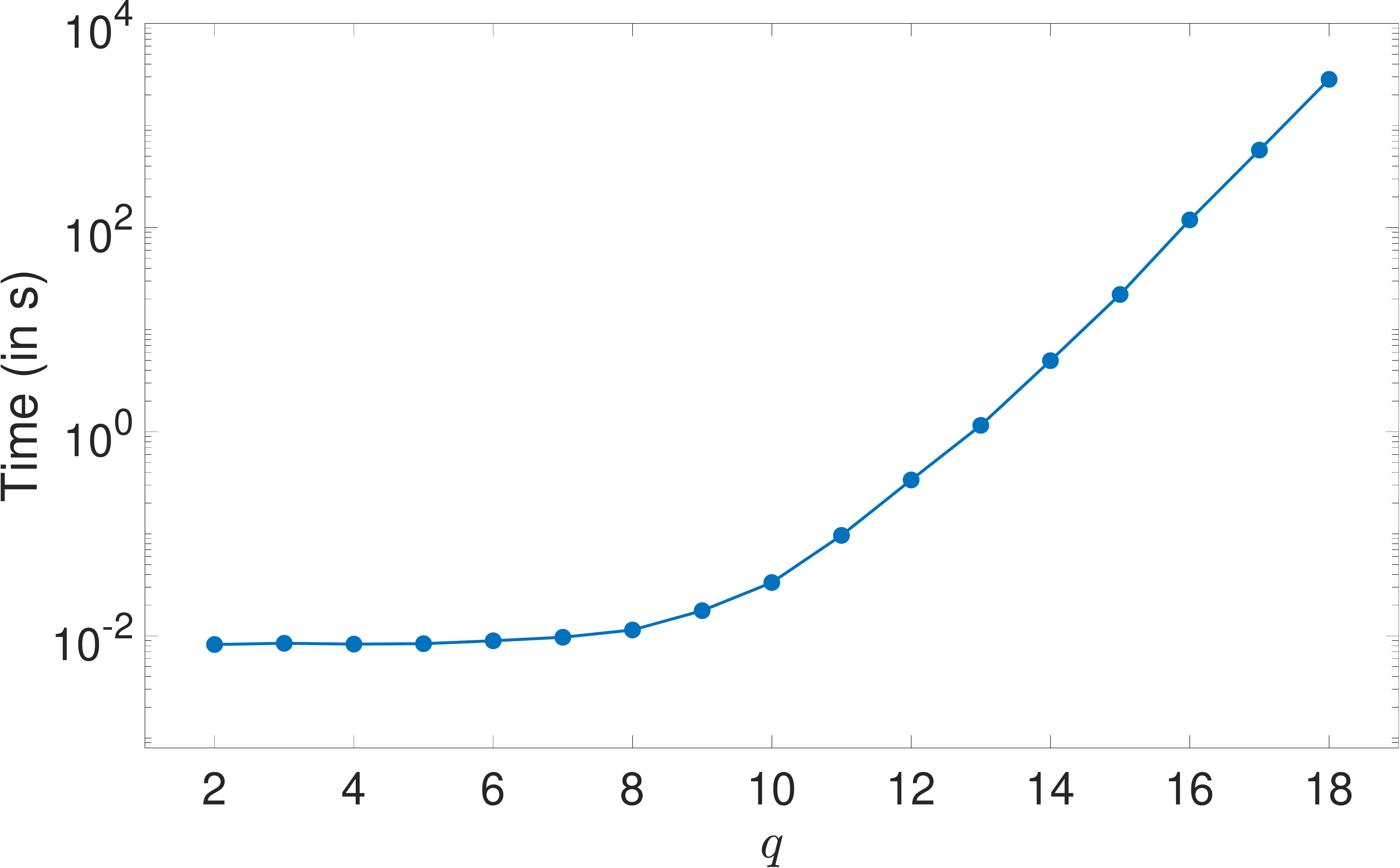}
	\caption{The computation time required to find a linear programming solution on a desktop computer as a function of $q$. For small values of $q$ this time is almost constant, but rises exponentially for larger $q$, rendering this method impractical beyond around 20 qubits. Timings are shown for the interior point algorithm, but results were very similar for the simplex algorithm. Error bars (estimated by repetition) are comparable to the size of the symbols, and the line simply joins the individual points.}
	\label{ctime}
\end{figure}

The principal downside of this approach is that the computational time needed to run linear programming increases with the size of the basis set, and this grows as $s=2^q$ for the current method. As discussed above a computation time scaling proportional to $(s+r)^2$ or $(s+r)^3$, where $r$ is the number of equality constraints, is likely, which for large $q$ is dominated by the exponential growth in $s$. This is confirmed by a plot of time required on a desktop computer (Intel Core i7-9700, 3.0--4.7\,GHz, with 12\,MB cache and 40\,GB RAM), shown in Fig.~\ref{ctime}. The linear behaviour at large $q$ on this semi-log plot indicates exponential computational time complexity, and the gradient is consistent with the time required scaling as about $4^q$.

Although attempts have been made to parallelize linear programming algorithms, progress so far has been limited \cite{Hall2010}. Thus, this method seems to be practical only up to a small number of spins, perhaps $q = 20$. Indeed much above 20 qubits it becomes difficult even to hold the $S$ matrix in memory on a desktop computer, although this could be sidestepped with a customised algorithm. While it is true that 20 is quite a large number of spins in the context of conventional NMR or even NMR QIP, we do not want to restrict ourselves to NMR spin systems but to extend to more general quantum systems which have the potential for a scalable quantum computing architecture. Fortunately, a more sophisticated approach is available which takes time only polynomial in $q$, albeit with a high power.

\section{Stabilizing solutions}
\label{stabilization}
Having located an optimal solution it is important to consider whether it can be implemented reliably. This will obviously require the ability to implement large numbers of $\pi$ pulses (single-qubit \textsc{not} gates) reliably, but we do not consider that problem here. Instead we consider the more fundamental question of the precision which is necessary for the free evolution times which lie between the pulses.

\begin{figure}[tb]
	\centering
	\includegraphics[width=85mm]{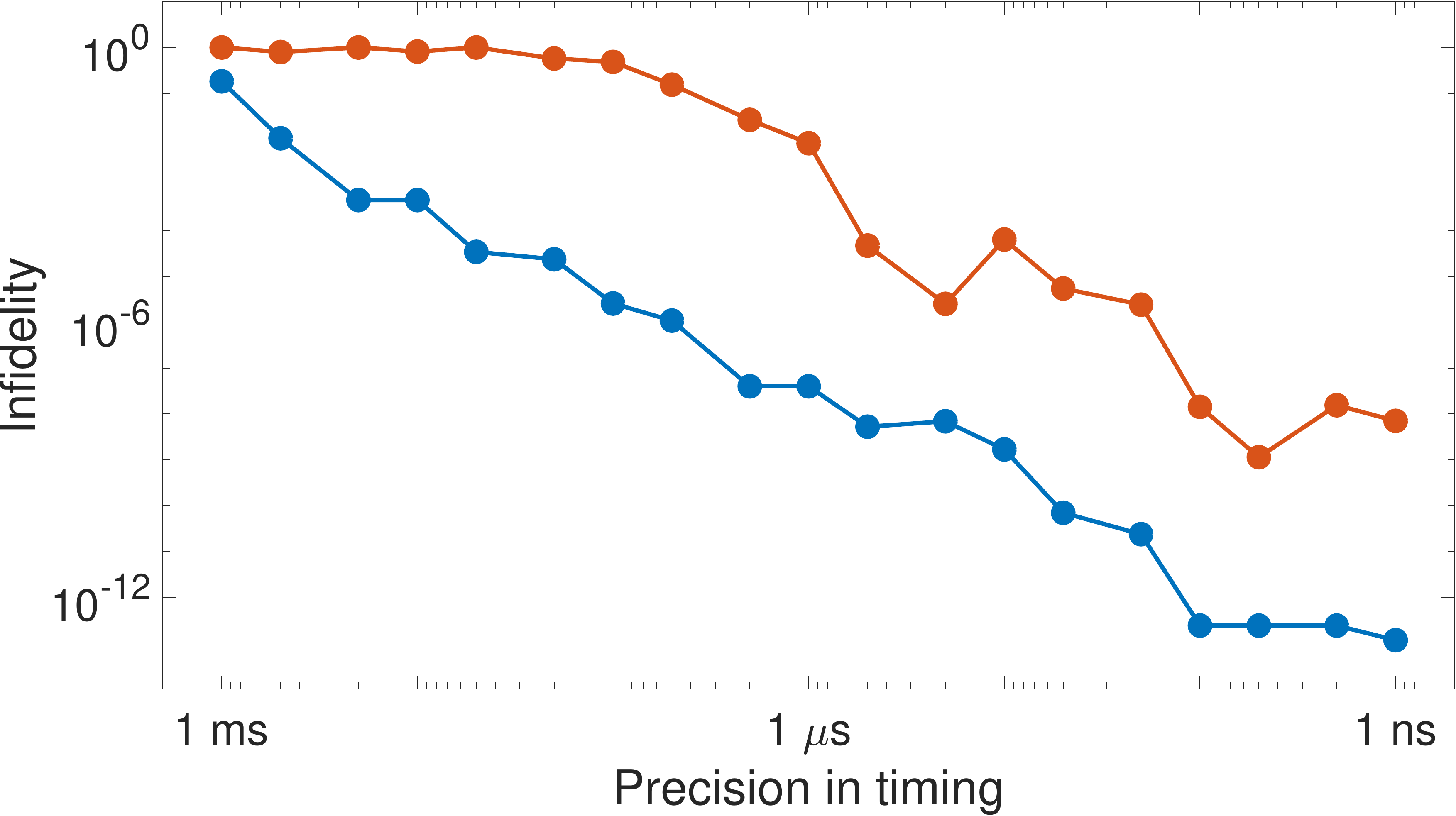}
	\caption{The infidelity of three-qubit circuits as a function of the precision of the underlying clock. Both the infidelity $\mathcal{I}=1-\mathcal{F}$ and the time resolution are plotted on logarithmic scales, with the time resolution running from 1\,ms to 1\,ns. The lower line, plotted in blue, shows the infidelity for the first three-qubit circuit, in which the single-qubit target phases are all zero. The upper line, plotted in red, corresponds to the second three-qubit circuit, where the single-qubit target phases are all $\pi$.}
	\label{Tround}
\end{figure}

So far we have assumed that these times can be implemented with effectively perfect precision, but any real implementation will be built around a clock, so that any delay period must be rounded to the nearest multiple of some underlying cycle time.  We show in Fig.~\ref{Tround} the effects of such rounding on the two three-qubit circuits considered in the previous section.
This figure shows the infidelity $\mathcal{I}=1-\mathcal{F}$, where $\mathcal{F}$, the fidelity between the desired propagator, $U$, and the propagator actually implemented, $V$, is the propagator fidelity \cite{Jones2003b}
\begin{equation}
\mathcal{F}=\left|\frac{\textrm{tr}(U^\dag V)}{\textrm{tr}(U^\dag U)}\right|^{2}.
\end{equation}
This is based on the Hilbert--Schmidt inner product between two unitary operators \cite{NCbook}, with the square-modulus value taken to remove global phase differences and the denominator acting to normalise the fidelity for the dimension of the underlying Hilbert space.  Many other fidelity definitions are in use \cite{Chuang1997,Bowdrey2002,Nielsen2002,Hofmann2005,Pedersen2007,Mayer2018}, some of which are more suitable for the experimental measurement of fidelities, but for comparing two unitary operations these are all closely related.

The lower line, plotted in blue, shows the infidelity for the first three-qubit circuit implemented on iodo\-tri\-fluoro\-ethene, in which the single-qubit target phases are all zero. Clearly this circuit works well, achieving an acceptable fidelity with a time resolution of $1\,\mu$s which is easily achievable in NMR experiments. The upper line, plotted in red, corresponds to the second three-qubit circuit, where the single-qubit target phases are all $\pi$. This circuit is unacceptably sensitive, requiring an unrealistic time resolution of about 1\,ns to reach an acceptable fidelity.

The sensitivity of the second circuit arises because it uses evolution at the one-spin frequencies to achieve one-qubit phases. In NMR one-spin frequencies (typically several kHz) are much larger than the coupling frequencies that generate two-qubit phases (typically tens of Hz), and so very small errors in evolution time will give large phase errors. (This will be less of a problem in other technologies where the qubits are individually addressed, as the offset frequency can in principle be chosen at will, and so be set near zero.) The first circuit overcomes this by using pairs of \textit{identical} time periods to cancel the single-spin evolution entirely, and when these identical time periods are rounded they remain identical, and so perfectly generate the desired single-qubit phase equal to 0.  If a non-zero single-qubit phase is required it can be generated in other ways, most simply by altering the relative phase of two $\pi$ pulses \cite{Morton2006a,Jones2013}.

\begin{figure}[tb]
	\centering
	\includegraphics[width=85mm]{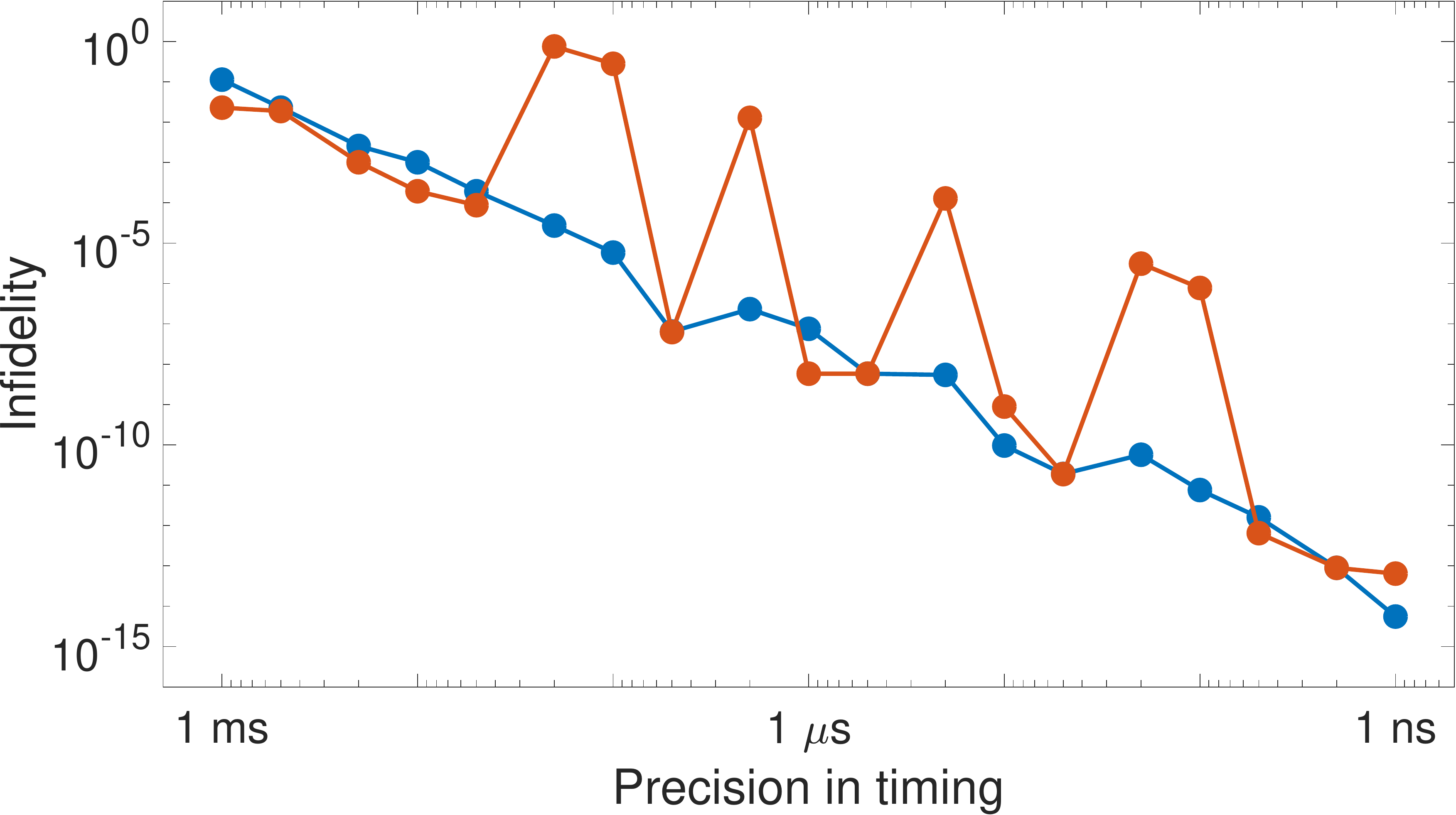}
	\caption{The infidelity of a four-qubit circuit (upper curve in red) and its stabilized equivalent (lower curve in blue) as a function of the precision of the underlying clock.}
	\label{Tround4}
\end{figure}

However, this restriction is not sufficient to guarantee a stable solution. The four-qubit network described in Eq.~\ref{R9} has all single-qubit phases equal to zero, and yet contains time periods without corresponding balancing pairs. The resulting infidelity is shown as the upper red curve in Fig.~\ref{Tround4}, and shows a very erratic dependence on time resolution.
Fortunately there is an easy method to stabilize such networks. If a reduced sign matrix $R$ is replaced by its negative $-R$ with the evolution times unchanged then this negated network will clearly generate identical two-spin phases to the original network (since the two-spin signs are negated twice, and so left unchanged) but the opposite one-spin phases. Thus if a matrix $R$ is combined with $-R$ and all the evolution times are halved the combined network will generate the original two-spin phases, but will have all one-spin phases equal to zero, even in the presence of rounding errors in time periods. The infidelity of the stablized four-qubit network is shown as the lower blue curve in Fig.~\ref{Tround4}, and shows a greatly improved dependence on time resolution.

By its construction the stabilized network will require the same total evolution time as the original network, and the same computational time to find it, and so the cost of stabilization is confined to a doubling in the number of time periods and $\pi$ pulses required for the implementation. This doubling can be reduced by subsequent optimization, locating equivalent evolutions in $R$ and $-R$. Of the nine sign combinations found in Eq.~\ref{R9}, six can also be found in $-R$, and so the corresponding evolution times can be combined. Thus the symmetrized circuit can be reduced from 18 periods down to only 12 distinct evolution periods.

As symmetrization is effectively essential to generate a stable network in the presence of large single-qubit interactions it is simpler just to generate symmetric networks by design. Since such networks are guaranteed to remove single-qubit phases, there are only $p=q(q-1)/2$ two-qubit phases that can be controlled. These phases can be controlled using only $p$ underlying times, each of which must occur twice, once in $R$ and once in $-R$. Thus it is unsurprising that a symmetrised four-qubit network contains twelve time periods. Such networks can be found by setting no constraints on the single-qubit phases, since these will be cancelled by the second half of the network. It is sufficient to use only columns from the first half of $S$ in the linear programming search, as columns from the second half will be used in $-R$. Thus the search for symmetrized networks is actually faster than for direct networks.

\section{Rescaling in larger spin systems}
The fundamental problem which slows down the rescaling algorithm is the exponentially large size of the overcomplete basis. This basis contains $s=2^q$ columns, from which the linear programming selects at most $r = q(q+1)/2$, equal to the number of interactions, and in many cases fewer.
For moderate values of $q$, these numbers $r$ and $s$ are quite similar, but the difference grows rapidly with $q$, leading one to wonder whether there is some way to cut down the size of $S$ before starting the linear programming step. Is it really necessary to include a very large number of columns, the great majority of which will eventually be discarded? In this regard we note that once one has identified the appropriate reduced sign matrix then the times required can be found by direct inversion of this $r$ by $r$ square matrix. However it is clear that most of the hard work is done in locating the appropriate columns used to construct the reduced sign matrix.

Starting from the other extreme, one could just select $r$ random columns from the full sign matrix and try to invert this. However, this process can fail in a number of ways. Firstly, the reduced matrix might not be of full rank (although this is unlikely when $s\gg r$) and so may not have an inverse. Even for a reduced matrix with full rank, the set of times obtained from the inversion process is very likely to include some negative times which are not physically implementable. Lastly, even if all the times are non-negative, the total time will not normally be the desired minimum, and so the sequence will not be time-optimal.

%\subsection{A random reduced overcomplete basis}
Between these extremes there is a middle way: using linear programming on a reduced, but still overcomplete, basis set. The current linear programming approach starts from the largest conceivable basis set, containing all of the $s=2^q$ possible sign patterns, which guarantees finding an optimal solution but also makes the process slow. One might imagine choosing some subset of columns at random, and attempting linear programming on this subset. For large values of $q$ the gap between the full size $s=2^q$ and the minimum size $r$ becomes very significant. It is thus worth exploring how many columns need to be picked so that linear programming generally finds a solution. There is no guarantee that such solutions will be time optimal, but as long as the random choice contains all the components of at least one optimal solution, then linear programming will find this. Given the very large number of equivalent solutions identified for moderate values of $q$ it seems plausible that this could be achieved with quite a small subset.

\subsection{The RROS method}
In the random reduced overcomplete set (RROS) method, instead of using all $s$ columns of the sign matrix $S$ as in the exhaustive approach, we choose just $kr$ columns from $S$, for some $k>1$, and run linear programming. Of course, one does not have to explicitly construct the entire $S$ matrix and then choose the $kr$ columns as these $kr$ columns correspond to the binary representations of $kr$ distinct decimal numbers chosen randomly from $0$ to $s-1$.  Our experience so far suggests that the probability of finding a possible solution, which achieves the desired phases using only positive evolution times, increases as $k$ increases, with a transition point around $k=2$, at which the probability of a random set giving a solution reaches 50\%.

\begin{figure}[tb]
	\centering
	\includegraphics[width=85mm]{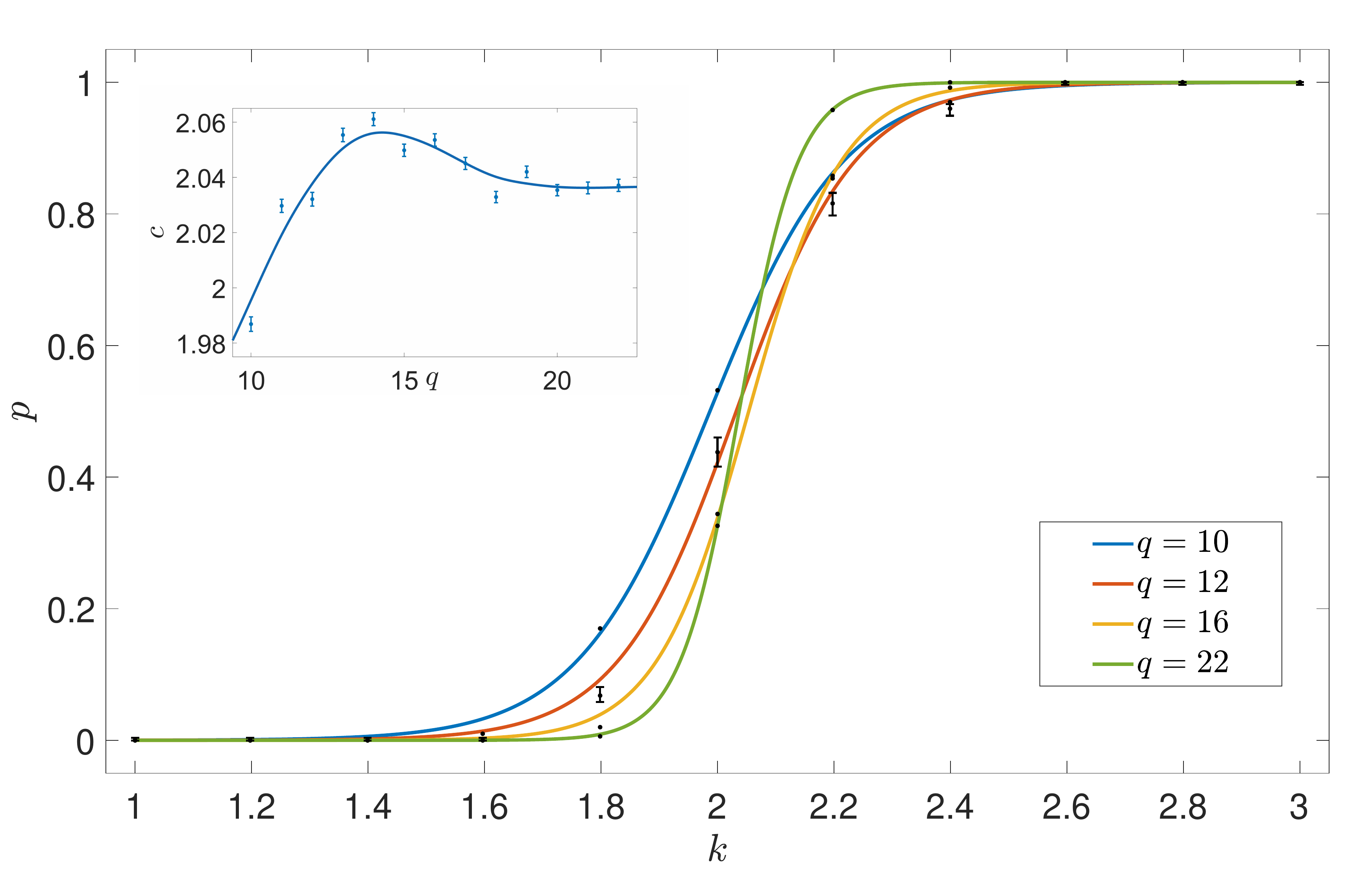}
	\caption{The effectiveness of the RROS method selecting $kr$ columns at random, showing how the probability of the algorithm finding a solution depends on $q$ and $k$. Calculations were performed for $q$ ranging from 10 to 22 but only four selected values are shown. Error bars on the $q=12$ points were estimated using Bayesian credible intervals; error bars for other curves are very similar.  Fitted sigmoidal curves were used to estimate the value of $k$ at which this success probability reached 50\%. The inset shows the  location of this transition point as a function of $q$. Error bars are estimated using error propagation from the sigmoidal fit results; the smooth curve has no significance and is plotted simply to guide the eye.}
	\label{pversusk}
\end{figure}

This observation is substantiated by the empirical evidence in Fig.~\ref{pversusk}.
RROS was run 500 times for values of $q$ ranging from 10 to 22 with $k$ varied between 1 and 3, and the fraction of occasions $f$ when linear programming found a suitable solution was calculated. Error bars on these estimates were calculated using Bayesian credible intervals \cite{Kruschke2011}, corresponding to the region of the probability density function within $\pm34\%$ of the median, equivalent to 1 standard deviation for a Normal distribution.  It is clear that the probability of success rises sharply as $k$ passes some critical value, with this transition becoming sharper as $q$ is increased. To help locate this transition point a sigmoidal logistic function \cite{Kruschke2011}
\begin{equation}
f(k)=\frac{1}{1+\exp[-b(x-c)]}
\end{equation}
was fitted for each value of $q$, with $c$ being the transition point at which the success probability passes 50\%, and $b$ indicating the sharpness of the transition. Although this function was chosen for convenience it clearly fits fairly well. A plot of the value of $c$ as a function of $q$ shown as an inset to Fig.~\ref{pversusk} suggests that the transition lies just above $k=2$.
For large values of $q$ this transition becomes sharp, so that for $k\ge2.5$ it is almost certain that a solution will be found.

If a solution is found then this solution will be time-optimal for the subset of columns chosen, but there is no guarantee that this will be the overall optimum, taking the shortest possible time.  Unsurprisingly the probability of finding a solution reaching the shortest possible time increases as $k$ increases, but investigating this question in detail is challenging, as for large values of $q$ the overall time-optimal solution cannot be located in a reasonable time. Nevertheless our preliminary studies suggest that for large $q$ the quality of solutions plateaus around $k\approx4$, and so there is little point going beyond this in practice. For small values of $q$ it seems to be necessary to use a slightly larger value of $k$, but in these cases it is more sensible just to use direct solution of the full $S$ matrix anyway.

\subsection{Results for large numbers of qubits}
\begin{figure}[tb]
	\centering
	\includegraphics[width=85mm]{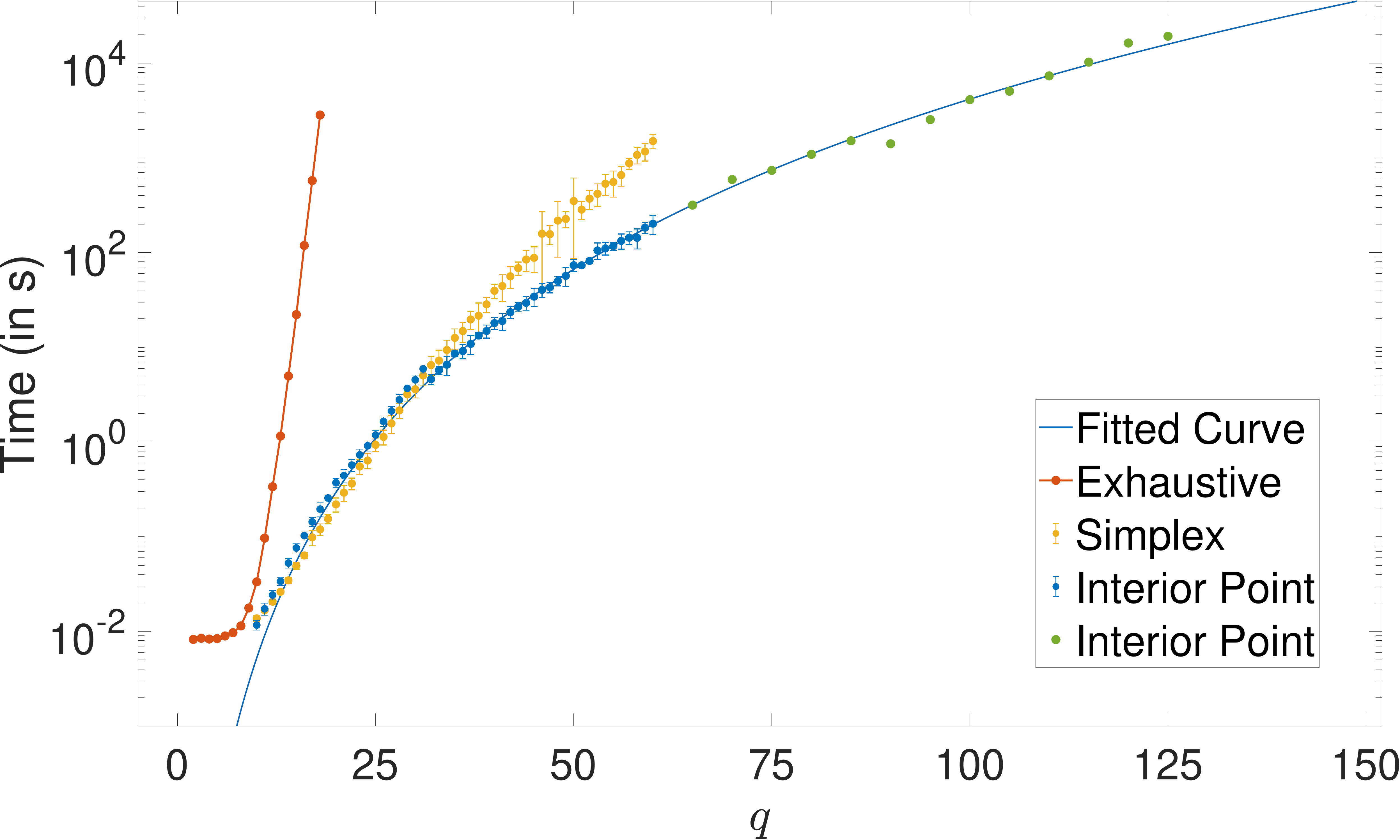}
	\caption{The computation time required to perform linear prediction using RROS with $k=4$. The simplex method was found to be fastest for $q\le31$, with the interior point method faster for $q\ge32$. Error bars for $q\le60$ indicated the mean and SD over 10 repetitions. The smooth curve with the form $Ar^3$ is fitted to interior point timings in the range $32\le q\le60$. Data points for $q>60$ are single repetitions and were not included in the fit. The fact that these points lie very close to the extrapolated fit suggests that our fitted curve is a good model. The red points on the left show timings for exhaustive calculations, demonstrating the huge time gains possible with RROS.}
	\label{RROST}
\end{figure}

The use of a smaller basis set permits RROS to be extended to much larger values of $q$. This was investigated by running the algorithm for $q$ between 10 and 60, as shown in Fig.~\ref{RROST}.
For RROS the time required to perform linear programming depends not only on the randomly chosen Hamiltonian and target phases but also on the precise choice of columns, and so timings were repeated 10 times using different choices. As expected for $k=4$ a solution was located in every case.

The simplex algorithm was found to be slightly faster than the interior point algorithm for $q\le31$, but the interior point algorithm was faster for $q\ge32$, and became much faster at high $q$. The discontinuity in the interior point timings between $q=31$ and $q=32$ may indicate a change in the precise algorithm used by Matlab.

For RROS the dimension of the problem is $n=r+kr$ and so we expect a computational complexity between $O(r^2)$ and $O(r^3)$. The smooth curve in the figure was fitted to the timings from the interior point algorithm for the range $32\le q\le60$ using a power function, and takes the form $A r^{3}$, consistent with our expectations (including the power as an additional fitting parameter gave a value indistinguishable from 3).   The timings for the simplex algorithm appear to lie between $O(r^4)$ and $O(r^5)$, which is much slower than the interior point.

With the simplex algorithm the number of time periods required was always approximately equal to $r$ (in some cases a small number of time periods were negligibly short and so could be dropped). The number of pulses required was determined for the $R$ matrix as originally found, without further optimisation, and was typically around $qr/2\approx q^3/4$, confirming our previous expectations.  With the interior point algorithm the solutions always use all $kr$ time periods, with a large number of very short times rather than a clear division into zero and non-zero times. These initial solutions can then be simplified by choosing a smaller subset of $k'r$ columns, with $k'\approx1.2$, corresponding to the columns with the largest time values in the original solution. Surprisingly the computational time  for these recalculations appears to be linear in $k$, while a quadratic or cubic dependence might have been expected.

The calculation was then extended to higher values of $q$, using only a single repetition, and these data points were found to lie very close to the extrapolated curve.  As $r\approx q^2/2$ this gives a pragmatic computational time scaling of $O(q^6)$, which is polynomial in the number of qubits.
Extrapolating this curve still further suggests that calculations with up to about 150 qubits could be practical, but the Matlab implementation \texttt{linprog} runs out of memory above $q=125$.  This limit should be solvable by a custom implementation of linear programming.

\subsection{Stability of solutions}
With large numbers of qubits it is not easy to directly evaluate the fidelity of a solution as even writing down the diagonal elements of the underlying propagator requires $2^q$ terms. It is, however, possible to estimate the infidelity directly using
\begin{equation}
\mathcal{I}\approx \sum_{i<j}\frac{(\delta\phi_{ij})^2}{16}
\end{equation}
where $\delta\phi_{ij}$ is the error in the coupling angle for the pair of qubits $i$ and $j$, and the approximation is only valid when the infidelity is small. This assumes that the solution has been symmetrised so that the errors in the single-spin angles are all zero.
Networks with large numbers of qubits have higher infidelities than those shown in section~\ref{stabilization}, reflecting the larger number of phases which need to be controlled. This is not a specific limitation of our methods, but is inherent in any attempt to control a large number of qubits simultaneously.

\section{Conclusions}
We have presented a practical algorithm which is guaranteed to find the minimum time solution to rescaling $z$ and $zz$ terms in the internal Hamiltonian of a quantum computer with up to about 20 qubits. Above 20 qubits the direct approach becomes intractable, but random sampling will extend this (although the solution might not be quite time optimal) to more than 100 qubits. Beyond about 150 qubits the $q^6$ scaling of the computational time for a fully coupled system renders any known approach impractical. It is, however, very unlikely that any system of that size would still be fully coupled, as couplings are usually only substantial between nearby spins. In  partially coupled systems $r$ is linear in $q$ rather than quadratic, and allowing for this should permit the process to be extended to hundreds or even thousands of qubits.

Although described in the language of NMR the techniques used are applicable to any equivalent system, where an always-on two-qubit interaction commutes with the single-qubit background terms. It may be of particular value in solid state platforms, such as superconducting circuits, in which 2D lattices of qubits are developed with a sparse coupling network (generally nearest-neighbour).  In such very large systems it is likely that the underlying structure in the pattern of interactions will allow symmetries to be exploited, potentially simplifying the problem greatly.

\begin{acknowledgments}
We thank Paul Goldberg (Department of Computer Science, University of Oxford) for useful discussions. GB is supported by a Felix Scholarship. TT is supported by the Masason Foundation and the Nakajima Foundation.
\end{acknowledgments}

\appendix
\section{Pulses for refocusing}
\label{app:pulsecount}
Consider a system of $q$ fully-coupled qubits where we seek to refocus all the one-qubit interactions and retain one single two-qubit interaction. This is most efficiently achieved by assigning the two coupled qubits to the Walsh function $W_1$, with the remaining qubits assigned from $W_2$ up to $W_{q-1}$ in sequence. As $W_n$ contains $n$ sign changes it requires $n$ $\pi$ pulses when $n$ is even, and $n+1$ pulses when $n$ is odd, with the additional pulse needed to restore the Hamiltonian to its original sign.

The overall number of pulse required is then obtained by summing these pulse counts from $n=1$ to $q-1$, remembering to include $W_1$ twice, giving a total of
\begin{equation*}
\frac{q(q-1)}{2}+\left\lceil\frac{q-1}{2}\right\rceil+2\le\frac{q^2}{2}+2.
\end{equation*}
Thus the total number of pulses required to retain a single two-qubit interaction is approximately $q^2/2$.
\section{Linear programming}
\label{app:lp}
The general linear programming problem seeks to maximize some linear function of its inputs subject to a set of linear equality and inequality constraints. In our application we seek to minimise the total evolution time, but this is easily achieved by maximising the negative of the total evolution, and our only inequality constraints are that all inputs (the individual evolution times for each period) must be positive.  The equality constraints then describe a convex polytope (a multidimensional polyhedron) called a simplex.

It can be shown that the solution to the linear programming problem always lies on one of the vertices, and so it is only necessary to compare the values of the function at all its vertices.  However for a high-dimensional problem the number of vertices becomes very large, and direct search is simply impractical.  Dantzig's original simplex algorithm provides a pragmatic search strategy that usually locates the optimum in a reasonable time.  Typically this time is no worse than $O(n^3)$, where $n$, the dimension of the problem, is in our case the sum of the number of inputs and the number of equality constraints. However it is known that particular problems can have worse scaling, and indeed some problems require full exhaustive comparison of every vertex.

A second common approach, the interior point algorithm, does not confine itself to vertices but instead travels across the inside of the polytope in an attempt to maximize the desired function while seeking not to cross the boundary faces of the polytope. This algorithm can be either faster or slower than Dantzig's algorithm, depending on details of the problem.

An important distinction between the two algorithms is that simplex will locate the true optimal vertex, and for our problem this has the property that some (typically the great majority) of the input times are strictly equal to zero, allowing these times to be dropped in any experimental implementation. By contrast the interior point algorithm only locates an approximate optimum, near one of the vertices, and so the solution contains a large number of input times which are small but not quite zero. These erroneous terms cannot simply be dropped without thought, as they act to cancel other error terms elsewhere in the solution. An important aspect of the stabilization method discussed in Section~\ref{stabilization} is that it enables these error terms to be dropped with comparative safety.
\bibliography{../../../all}

\begin{thebibliography}{51}
\expandafter\ifx\csname natexlab\endcsname\relax\def\natexlab#1{#1}\fi
\expandafter\ifx\csname bibnamefont\endcsname\relax
  \def\bibnamefont#1{#1}\fi
\expandafter\ifx\csname bibfnamefont\endcsname\relax
  \def\bibfnamefont#1{#1}\fi
\expandafter\ifx\csname citenamefont\endcsname\relax
  \def\citenamefont#1{#1}\fi
\expandafter\ifx\csname url\endcsname\relax
  \def\url#1{\texttt{#1}}\fi
\expandafter\ifx\csname urlprefix\endcsname\relax\def\urlprefix{URL }\fi
\providecommand{\bibinfo}[2]{#2}
\providecommand{\eprint}[2][]{\url{#2}}

\bibitem[{\citenamefont{Brown et~al.}(2016)\citenamefont{Brown, Kim, and
  Monroe}}]{Brown2016}
\bibinfo{author}{\bibfnamefont{K.~R.} \bibnamefont{Brown}},
  \bibinfo{author}{\bibfnamefont{J.}~\bibnamefont{Kim}}, \bibnamefont{and}
  \bibinfo{author}{\bibfnamefont{C.}~\bibnamefont{Monroe}},
  \bibinfo{journal}{npj Quantum Information} \textbf{\bibinfo{volume}{2}},
  \bibinfo{pages}{16034} (\bibinfo{year}{2016}).

\bibitem[{\citenamefont{Sch\"afer et~al.}(2018)\citenamefont{Sch\"afer,
  Ballance, Thirumalai, Stephenson, Ballance, Steane, and
  Lucas}}]{Schaefer2018}
\bibinfo{author}{\bibfnamefont{V.~M.} \bibnamefont{Sch\"afer}},
  \bibinfo{author}{\bibfnamefont{C.~J.} \bibnamefont{Ballance}},
  \bibinfo{author}{\bibfnamefont{K.}~\bibnamefont{Thirumalai}},
  \bibinfo{author}{\bibfnamefont{L.~J.} \bibnamefont{Stephenson}},
  \bibinfo{author}{\bibfnamefont{T.~G.} \bibnamefont{Ballance}},
  \bibinfo{author}{\bibfnamefont{A.~M.} \bibnamefont{Steane}},
  \bibnamefont{and} \bibinfo{author}{\bibfnamefont{D.~M.} \bibnamefont{Lucas}},
  \bibinfo{journal}{Nature} \textbf{\bibinfo{volume}{555}}, \bibinfo{pages}{75}
  (\bibinfo{year}{2018}).

\bibitem[{\citenamefont{Jones}(2011)}]{Jones2011}
\bibinfo{author}{\bibfnamefont{J.~A.} \bibnamefont{Jones}},
  \bibinfo{journal}{Prog. NMR Spectrosc.} \textbf{\bibinfo{volume}{59}},
  \bibinfo{pages}{91} (\bibinfo{year}{2011}).

\bibitem[{\citenamefont{Barenco et~al.}(1995)\citenamefont{Barenco, Bennett,
  Cleve, DiVincenzo, Margolus, Shor, Sleator, Smolin, and
  Weinfurter}}]{Barenco1995}
\bibinfo{author}{\bibfnamefont{A.}~\bibnamefont{Barenco}},
  \bibinfo{author}{\bibfnamefont{C.~H.} \bibnamefont{Bennett}},
  \bibinfo{author}{\bibfnamefont{R.}~\bibnamefont{Cleve}},
  \bibinfo{author}{\bibfnamefont{D.~P.} \bibnamefont{DiVincenzo}},
  \bibinfo{author}{\bibfnamefont{N.}~\bibnamefont{Margolus}},
  \bibinfo{author}{\bibfnamefont{P.}~\bibnamefont{Shor}},
  \bibinfo{author}{\bibfnamefont{T.}~\bibnamefont{Sleator}},
  \bibinfo{author}{\bibfnamefont{J.~A.} \bibnamefont{Smolin}},
  \bibnamefont{and}
  \bibinfo{author}{\bibfnamefont{H.}~\bibnamefont{Weinfurter}},
  \bibinfo{journal}{Phys. Rev. A} \textbf{\bibinfo{volume}{52}},
  \bibinfo{pages}{3457} (\bibinfo{year}{1995}).

\bibitem[{\citenamefont{Kane}(1998)}]{Kane1998}
\bibinfo{author}{\bibfnamefont{B.~E.} \bibnamefont{Kane}},
  \bibinfo{journal}{Nature} \textbf{\bibinfo{volume}{393}},
  \bibinfo{pages}{133} (\bibinfo{year}{1998}).

\bibitem[{\citenamefont{Loss and DiVincenzo}(1998)}]{Loss1998}
\bibinfo{author}{\bibfnamefont{D.}~\bibnamefont{Loss}} \bibnamefont{and}
  \bibinfo{author}{\bibfnamefont{D.~P.} \bibnamefont{DiVincenzo}},
  \bibinfo{journal}{Phys. Rev. A} \textbf{\bibinfo{volume}{57}},
  \bibinfo{pages}{120} (\bibinfo{year}{1998}).

\bibitem[{\citenamefont{Brennen et~al.}(1999)\citenamefont{Brennen, Caves,
  Jessen, and Deutsch}}]{Brennen1999}
\bibinfo{author}{\bibfnamefont{G.~K.} \bibnamefont{Brennen}},
  \bibinfo{author}{\bibfnamefont{C.~M.} \bibnamefont{Caves}},
  \bibinfo{author}{\bibfnamefont{P.~S.} \bibnamefont{Jessen}},
  \bibnamefont{and} \bibinfo{author}{\bibfnamefont{I.~H.}
  \bibnamefont{Deutsch}}, \bibinfo{journal}{Phys. Rev. Lett.}
  \textbf{\bibinfo{volume}{82}}, \bibinfo{pages}{1060} (\bibinfo{year}{1999}).

\bibitem[{\citenamefont{Jaksch et~al.}(1999)\citenamefont{Jaksch, Briegel,
  Cirac, Gardiner, and Zoller}}]{Jaksch1999}
\bibinfo{author}{\bibfnamefont{D.}~\bibnamefont{Jaksch}},
  \bibinfo{author}{\bibfnamefont{H.-J.} \bibnamefont{Briegel}},
  \bibinfo{author}{\bibfnamefont{J.~I.} \bibnamefont{Cirac}},
  \bibinfo{author}{\bibfnamefont{C.~W.} \bibnamefont{Gardiner}},
  \bibnamefont{and} \bibinfo{author}{\bibfnamefont{P.}~\bibnamefont{Zoller}},
  \bibinfo{journal}{Phys. Rev. Lett.} \textbf{\bibinfo{volume}{82}},
  \bibinfo{pages}{1975} (\bibinfo{year}{1999}).

\bibitem[{\citenamefont{Knill et~al.}(2001)\citenamefont{Knill, Laflamme,
  Martinez, and Negrevergne}}]{Knill2001}
\bibinfo{author}{\bibfnamefont{E.}~\bibnamefont{Knill}},
  \bibinfo{author}{\bibfnamefont{R.}~\bibnamefont{Laflamme}},
  \bibinfo{author}{\bibfnamefont{R.}~\bibnamefont{Martinez}}, \bibnamefont{and}
  \bibinfo{author}{\bibfnamefont{C.}~\bibnamefont{Negrevergne}},
  \bibinfo{journal}{Phys. Rev. Lett.} \textbf{\bibinfo{volume}{86}},
  \bibinfo{pages}{5811} (\bibinfo{year}{2001}).

\bibitem[{\citenamefont{Leuenberger and Loss}(2001)}]{Leuenberger2001}
\bibinfo{author}{\bibfnamefont{M.~N.} \bibnamefont{Leuenberger}}
  \bibnamefont{and} \bibinfo{author}{\bibfnamefont{D.}~\bibnamefont{Loss}},
  \bibinfo{journal}{Nature} \textbf{\bibinfo{volume}{410}},
  \bibinfo{pages}{789} (\bibinfo{year}{2001}).

\bibitem[{\citenamefont{Harneit}(2002)}]{Harneit2002}
\bibinfo{author}{\bibfnamefont{W.}~\bibnamefont{Harneit}},
  \bibinfo{journal}{Phys. Rev. A} \textbf{\bibinfo{volume}{65}},
  \bibinfo{pages}{032322} (\bibinfo{year}{2002}).

\bibitem[{\citenamefont{Suter and Lim}(2002)}]{Suter2002}
\bibinfo{author}{\bibfnamefont{D.}~\bibnamefont{Suter}} \bibnamefont{and}
  \bibinfo{author}{\bibfnamefont{K.}~\bibnamefont{Lim}},
  \bibinfo{journal}{Phys. Rev. A} \textbf{\bibinfo{volume}{65}},
  \bibinfo{pages}{052309} (\bibinfo{year}{2002}).

\bibitem[{\citenamefont{Chen et~al.}(2019)\citenamefont{Chen, Zhou, and
  Shen}}]{Chen2019}
\bibinfo{author}{\bibfnamefont{Z.}~\bibnamefont{Chen}},
  \bibinfo{author}{\bibfnamefont{Y.}~\bibnamefont{Zhou}}, \bibnamefont{and}
  \bibinfo{author}{\bibfnamefont{J.-T.} \bibnamefont{Shen}}, in
  \emph{\bibinfo{booktitle}{Conference on Lasers and Electro-Optics}}
  (\bibinfo{publisher}{Optical Society of America}, \bibinfo{year}{2019}), p.
  \bibinfo{pages}{FTh3A.4}.

\bibitem[{\citenamefont{Parra-Rodriguez
  et~al.}(2018)\citenamefont{Parra-Rodriguez, Lougovski, Lamata, Solano, and
  Sanz}}]{Parra-Rodriguez2018}
\bibinfo{author}{\bibfnamefont{A.}~\bibnamefont{Parra-Rodriguez}},
  \bibinfo{author}{\bibfnamefont{P.}~\bibnamefont{Lougovski}},
  \bibinfo{author}{\bibfnamefont{L.}~\bibnamefont{Lamata}},
  \bibinfo{author}{\bibfnamefont{E.}~\bibnamefont{Solano}}, \bibnamefont{and}
  \bibinfo{author}{\bibfnamefont{M.}~\bibnamefont{Sanz}},
  \bibinfo{journal}{Arxiv preprint arXiv:1812.03637}  (\bibinfo{year}{2018}).

\bibitem[{\citenamefont{Wu et~al.}(2019)\citenamefont{Wu, Wang, Qin, Rong, and
  Du}}]{Wu2019}
\bibinfo{author}{\bibfnamefont{Y.}~\bibnamefont{Wu}},
  \bibinfo{author}{\bibfnamefont{Y.}~\bibnamefont{Wang}},
  \bibinfo{author}{\bibfnamefont{X.}~\bibnamefont{Qin}},
  \bibinfo{author}{\bibfnamefont{X.}~\bibnamefont{Rong}}, \bibnamefont{and}
  \bibinfo{author}{\bibfnamefont{J.}~\bibnamefont{Du}}, \bibinfo{journal}{npj
  Quantum Information} \textbf{\bibinfo{volume}{5}}, \bibinfo{pages}{9}
  (\bibinfo{year}{2019}).

\bibitem[{\citenamefont{Moehring et~al.}(2007)\citenamefont{Moehring, Maunz,
  Olmschenk, Younge, Matsukevich, Duan, and Monroe}}]{Moehring2007}
\bibinfo{author}{\bibfnamefont{D.~L.} \bibnamefont{Moehring}},
  \bibinfo{author}{\bibfnamefont{P.}~\bibnamefont{Maunz}},
  \bibinfo{author}{\bibfnamefont{S.}~\bibnamefont{Olmschenk}},
  \bibinfo{author}{\bibfnamefont{K.~C.} \bibnamefont{Younge}},
  \bibinfo{author}{\bibfnamefont{D.~N.} \bibnamefont{Matsukevich}},
  \bibinfo{author}{\bibfnamefont{L.-M.} \bibnamefont{Duan}}, \bibnamefont{and}
  \bibinfo{author}{\bibfnamefont{C.}~\bibnamefont{Monroe}},
  \bibinfo{journal}{Nature} \textbf{\bibinfo{volume}{449}}, \bibinfo{pages}{68}
  (\bibinfo{year}{2007}).

\bibitem[{\citenamefont{Arute et~al.}(2019)\citenamefont{Arute, Arya, Babbush,
  Bacon, Bardin, Barends, Biswas, Boixo, Brandao, Buell et~al.}}]{Arute2019}
\bibinfo{author}{\bibfnamefont{F.}~\bibnamefont{Arute}},
  \bibinfo{author}{\bibfnamefont{K.}~\bibnamefont{Arya}},
  \bibinfo{author}{\bibfnamefont{R.}~\bibnamefont{Babbush}},
  \bibinfo{author}{\bibfnamefont{D.}~\bibnamefont{Bacon}},
  \bibinfo{author}{\bibfnamefont{J.~C.} \bibnamefont{Bardin}},
  \bibinfo{author}{\bibfnamefont{R.}~\bibnamefont{Barends}},
  \bibinfo{author}{\bibfnamefont{R.}~\bibnamefont{Biswas}},
  \bibinfo{author}{\bibfnamefont{S.}~\bibnamefont{Boixo}},
  \bibinfo{author}{\bibfnamefont{F.~G. S.~L.} \bibnamefont{Brandao}},
  \bibinfo{author}{\bibfnamefont{D.~A.} \bibnamefont{Buell}},
  \bibnamefont{et~al.}, \bibinfo{journal}{Nature}
  \textbf{\bibinfo{volume}{574}}, \bibinfo{pages}{505} (\bibinfo{year}{2019}).

\bibitem[{\citenamefont{Gambetta et~al.}(2017)\citenamefont{Gambetta, Chow, and
  Steffen}}]{Gambetta2017}
\bibinfo{author}{\bibfnamefont{J.~M.} \bibnamefont{Gambetta}},
  \bibinfo{author}{\bibfnamefont{J.~M.} \bibnamefont{Chow}}, \bibnamefont{and}
  \bibinfo{author}{\bibfnamefont{M.}~\bibnamefont{Steffen}},
  \bibinfo{journal}{npj Quantum Information} \textbf{\bibinfo{volume}{3}},
  \bibinfo{pages}{2} (\bibinfo{year}{2017}).

\bibitem[{\citenamefont{Hahn}(1950)}]{Hahn1950}
\bibinfo{author}{\bibfnamefont{E.~L.} \bibnamefont{Hahn}},
  \bibinfo{journal}{Phys. Rev.} \textbf{\bibinfo{volume}{80}},
  \bibinfo{pages}{580} (\bibinfo{year}{1950}).

\bibitem[{\citenamefont{Freeman}(1997)}]{FreemanSCbook}
\bibinfo{author}{\bibfnamefont{R.}~\bibnamefont{Freeman}},
  \emph{\bibinfo{title}{Spin Choreography}} (\bibinfo{publisher}{Spektrum},
  \bibinfo{year}{1997}).

\bibitem[{\citenamefont{Jones and Knill}(1999)}]{Jones1999}
\bibinfo{author}{\bibfnamefont{J.~A.} \bibnamefont{Jones}} \bibnamefont{and}
  \bibinfo{author}{\bibfnamefont{E.}~\bibnamefont{Knill}}, \bibinfo{journal}{J.
  Magn. Reson.} \textbf{\bibinfo{volume}{141}}, \bibinfo{pages}{322}
  (\bibinfo{year}{1999}).

\bibitem[{\citenamefont{Leung et~al.}(2000)\citenamefont{Leung, Chuang,
  Yamaguchi, and Yamamoto}}]{Leung2000}
\bibinfo{author}{\bibfnamefont{D.~W.} \bibnamefont{Leung}},
  \bibinfo{author}{\bibfnamefont{I.~L.} \bibnamefont{Chuang}},
  \bibinfo{author}{\bibfnamefont{F.}~\bibnamefont{Yamaguchi}},
  \bibnamefont{and} \bibinfo{author}{\bibfnamefont{Y.}~\bibnamefont{Yamamoto}},
  \bibinfo{journal}{Phys. Rev. A} \textbf{\bibinfo{volume}{61}},
  \bibinfo{pages}{42310} (\bibinfo{year}{2000}).

\bibitem[{\citenamefont{Leung}(2002)}]{Leung2002}
\bibinfo{author}{\bibfnamefont{D.}~\bibnamefont{Leung}}, \bibinfo{journal}{J.
  Mod. Opt.} \textbf{\bibinfo{volume}{49}}, \bibinfo{pages}{1199}
  (\bibinfo{year}{2002}).

\bibitem[{\citenamefont{Hayes et~al.}(2014)\citenamefont{Hayes, Flammia, and
  Biercuk}}]{Hayes2014}
\bibinfo{author}{\bibfnamefont{D.}~\bibnamefont{Hayes}},
  \bibinfo{author}{\bibfnamefont{S.~T.} \bibnamefont{Flammia}},
  \bibnamefont{and} \bibinfo{author}{\bibfnamefont{M.~J.}
  \bibnamefont{Biercuk}}, \bibinfo{journal}{New J. Phys.}
  \textbf{\bibinfo{volume}{16}}, \bibinfo{pages}{083027}
  (\bibinfo{year}{2014}).

\bibitem[{\citenamefont{Welch et~al.}(2014)\citenamefont{Welch, Greenbaum,
  Mostame, and Aspuru-Guzik}}]{Welch2014}
\bibinfo{author}{\bibfnamefont{J.}~\bibnamefont{Welch}},
  \bibinfo{author}{\bibfnamefont{D.}~\bibnamefont{Greenbaum}},
  \bibinfo{author}{\bibfnamefont{S.}~\bibnamefont{Mostame}}, \bibnamefont{and}
  \bibinfo{author}{\bibfnamefont{A.}~\bibnamefont{Aspuru-Guzik}},
  \bibinfo{journal}{New J. Phys.} \textbf{\bibinfo{volume}{16}},
  \bibinfo{pages}{033040} (\bibinfo{year}{2014}).

\bibitem[{\citenamefont{Beauchamp}(1984)}]{Beauchamp1984}
\bibinfo{author}{\bibfnamefont{K.~G.} \bibnamefont{Beauchamp}},
  \emph{\bibinfo{title}{Applications of Walsh and related functions}}
  (\bibinfo{publisher}{Academic Press}, \bibinfo{year}{1984}).

\bibitem[{\citenamefont{Lynn}(1964)}]{Lynn1964}
\bibinfo{author}{\bibfnamefont{M.~S.} \bibnamefont{Lynn}},
  \bibinfo{journal}{Math. Proc. Camb. Phil. Soc.}
  \textbf{\bibinfo{volume}{60}}, \bibinfo{pages}{425} (\bibinfo{year}{1964}).

\bibitem[{\citenamefont{Bland}(1981)}]{Bland1981}
\bibinfo{author}{\bibfnamefont{R.~G.} \bibnamefont{Bland}},
  \bibinfo{journal}{Scientific American} \textbf{\bibinfo{volume}{244}},
  \bibinfo{pages}{126} (\bibinfo{year}{1981}).

\bibitem[{\citenamefont{Dantzig}(1982)}]{Dantzig1982}
\bibinfo{author}{\bibfnamefont{G.~B.} \bibnamefont{Dantzig}},
  \bibinfo{journal}{Operations Research Letters} \textbf{\bibinfo{volume}{1}},
  \bibinfo{pages}{43 } (\bibinfo{year}{1982}).

\bibitem[{\citenamefont{Adler et~al.}(1989)\citenamefont{Adler, Resende, Veiga,
  and Karmarkar}}]{Adler1989}
\bibinfo{author}{\bibfnamefont{I.}~\bibnamefont{Adler}},
  \bibinfo{author}{\bibfnamefont{M.}~\bibnamefont{Resende}},
  \bibinfo{author}{\bibfnamefont{G.}~\bibnamefont{Veiga}}, \bibnamefont{and}
  \bibinfo{author}{\bibfnamefont{N.}~\bibnamefont{Karmarkar}},
  \bibinfo{journal}{Mathematical Programming} \textbf{\bibinfo{volume}{44}},
  \bibinfo{pages}{297} (\bibinfo{year}{1989}).

\bibitem[{\citenamefont{Press et~al.}(1992)\citenamefont{Press, Teukolsky,
  Vettering, and Flannery}}]{NumRec1992}
\bibinfo{author}{\bibfnamefont{W.~H.} \bibnamefont{Press}},
  \bibinfo{author}{\bibfnamefont{S.~A.} \bibnamefont{Teukolsky}},
  \bibinfo{author}{\bibfnamefont{W.~T.} \bibnamefont{Vettering}},
  \bibnamefont{and} \bibinfo{author}{\bibfnamefont{B.~P.}
  \bibnamefont{Flannery}}, \emph{\bibinfo{title}{Numerical Recipes in C}}
  (\bibinfo{publisher}{Cambridge University Press}, \bibinfo{year}{1992}),
  \bibinfo{edition}{2nd} ed.

\bibitem[{\citenamefont{Du et~al.}(2007)\citenamefont{Du, Zhu, Shi, Peng, and
  Suter}}]{Du2007}
\bibinfo{author}{\bibfnamefont{J.}~\bibnamefont{Du}},
  \bibinfo{author}{\bibfnamefont{J.}~\bibnamefont{Zhu}},
  \bibinfo{author}{\bibfnamefont{M.}~\bibnamefont{Shi}},
  \bibinfo{author}{\bibfnamefont{X.}~\bibnamefont{Peng}}, \bibnamefont{and}
  \bibinfo{author}{\bibfnamefont{D.}~\bibnamefont{Suter}},
  \bibinfo{journal}{Phys. Rev. A} \textbf{\bibinfo{volume}{76}},
  \bibinfo{pages}{042121} (\bibinfo{year}{2007}).

\bibitem[{\citenamefont{Knill et~al.}(2000)\citenamefont{Knill, Laflamme,
  Martinez, and Tseng}}]{Knill2000}
\bibinfo{author}{\bibfnamefont{E.}~\bibnamefont{Knill}},
  \bibinfo{author}{\bibfnamefont{R.}~\bibnamefont{Laflamme}},
  \bibinfo{author}{\bibfnamefont{R.}~\bibnamefont{Martinez}}, \bibnamefont{and}
  \bibinfo{author}{\bibfnamefont{C.~H.} \bibnamefont{Tseng}},
  \bibinfo{journal}{Nature} \textbf{\bibinfo{volume}{404}},
  \bibinfo{pages}{368} (\bibinfo{year}{2000}).

\bibitem[{\citenamefont{Bowdrey et~al.}(2005)\citenamefont{Bowdrey, Jones,
  Knill, and Laflamme}}]{Bowdrey2005}
\bibinfo{author}{\bibfnamefont{M.~D.} \bibnamefont{Bowdrey}},
  \bibinfo{author}{\bibfnamefont{J.~A.} \bibnamefont{Jones}},
  \bibinfo{author}{\bibfnamefont{E.}~\bibnamefont{Knill}}, \bibnamefont{and}
  \bibinfo{author}{\bibfnamefont{R.}~\bibnamefont{Laflamme}},
  \bibinfo{journal}{Phys. Rev. A} \textbf{\bibinfo{volume}{72}},
  \bibinfo{pages}{032315} (\bibinfo{year}{2005}).

\bibitem[{\citenamefont{Freeman et~al.}(1981)\citenamefont{Freeman, Frenkiel,
  and Levitt}}]{Freeman1981}
\bibinfo{author}{\bibfnamefont{R.}~\bibnamefont{Freeman}},
  \bibinfo{author}{\bibfnamefont{T.~A.} \bibnamefont{Frenkiel}},
  \bibnamefont{and} \bibinfo{author}{\bibfnamefont{M.~H.}
  \bibnamefont{Levitt}}, \bibinfo{journal}{J. Magn. Reson}
  \textbf{\bibinfo{volume}{44}}, \bibinfo{pages}{409} (\bibinfo{year}{1981}).

\bibitem[{\citenamefont{Morton et~al.}(2006)\citenamefont{Morton, Tyryshkin,
  Ardavan, Benjamin, Porfyrakis, Lyon, and Briggs}}]{Morton2006a}
\bibinfo{author}{\bibfnamefont{J.~J.~L.} \bibnamefont{Morton}},
  \bibinfo{author}{\bibfnamefont{A.~M.} \bibnamefont{Tyryshkin}},
  \bibinfo{author}{\bibfnamefont{A.}~\bibnamefont{Ardavan}},
  \bibinfo{author}{\bibfnamefont{S.~C.} \bibnamefont{Benjamin}},
  \bibinfo{author}{\bibfnamefont{K.}~\bibnamefont{Porfyrakis}},
  \bibinfo{author}{\bibfnamefont{S.~A.} \bibnamefont{Lyon}}, \bibnamefont{and}
  \bibinfo{author}{\bibfnamefont{G.~A.~D.} \bibnamefont{Briggs}},
  \bibinfo{journal}{Nature Physics} \textbf{\bibinfo{volume}{2}},
  \bibinfo{pages}{40} (\bibinfo{year}{2006}).

\bibitem[{\citenamefont{Jones}(2013)}]{Jones2013}
\bibinfo{author}{\bibfnamefont{J.~A.} \bibnamefont{Jones}},
  \bibinfo{journal}{Phys. Rev. A} \textbf{\bibinfo{volume}{87}},
  \bibinfo{pages}{052317} (\bibinfo{year}{2013}).

\bibitem[{\citenamefont{Khaneja et~al.}(2005)\citenamefont{Khaneja, Reiss,
  Kehlet, Schulte-Herbr{\"u}ggen, and Glaser}}]{Khaneja2005}
\bibinfo{author}{\bibfnamefont{N.}~\bibnamefont{Khaneja}},
  \bibinfo{author}{\bibfnamefont{T.}~\bibnamefont{Reiss}},
  \bibinfo{author}{\bibfnamefont{C.}~\bibnamefont{Kehlet}},
  \bibinfo{author}{\bibfnamefont{T.}~\bibnamefont{Schulte-Herbr{\"u}ggen}},
  \bibnamefont{and} \bibinfo{author}{\bibfnamefont{S.~J.}
  \bibnamefont{Glaser}}, \bibinfo{journal}{J. Magn. Reson.}
  \textbf{\bibinfo{volume}{172}}, \bibinfo{pages}{296} (\bibinfo{year}{2005}).

\bibitem[{\citenamefont{Boulant et~al.}(2002)\citenamefont{Boulant, Fortunato,
  Pravia, Teklemariam, Cory, and Havel}}]{Boulant2002}
\bibinfo{author}{\bibfnamefont{N.}~\bibnamefont{Boulant}},
  \bibinfo{author}{\bibfnamefont{E.~M.} \bibnamefont{Fortunato}},
  \bibinfo{author}{\bibfnamefont{M.~A.} \bibnamefont{Pravia}},
  \bibinfo{author}{\bibfnamefont{G.}~\bibnamefont{Teklemariam}},
  \bibinfo{author}{\bibfnamefont{D.~G.} \bibnamefont{Cory}}, \bibnamefont{and}
  \bibinfo{author}{\bibfnamefont{T.~F.} \bibnamefont{Havel}},
  \bibinfo{journal}{Phys. Rev. A} \textbf{\bibinfo{volume}{65}},
  \bibinfo{pages}{024302} (\bibinfo{year}{2002}).

\bibitem[{\citenamefont{Collins et~al.}(2000)\citenamefont{Collins, Kim,
  Holton, Sierzputowska-Gracz, and Stejskal}}]{Collins2000}
\bibinfo{author}{\bibfnamefont{D.}~\bibnamefont{Collins}},
  \bibinfo{author}{\bibfnamefont{K.~W.} \bibnamefont{Kim}},
  \bibinfo{author}{\bibfnamefont{W.~C.} \bibnamefont{Holton}},
  \bibinfo{author}{\bibfnamefont{H.}~\bibnamefont{Sierzputowska-Gracz}},
  \bibnamefont{and} \bibinfo{author}{\bibfnamefont{E.~O.}
  \bibnamefont{Stejskal}}, \bibinfo{journal}{Phys. Rev. A}
  \textbf{\bibinfo{volume}{62}}, \bibinfo{pages}{022304}
  (\bibinfo{year}{2000}).

\bibitem[{\citenamefont{Khaneja et~al.}(2007)\citenamefont{Khaneja, Heitmann,
  Sp{\"o}rl, Yuan, Schulte-Herbr{\"u}ggen, and Glaser}}]{Khaneja2007}
\bibinfo{author}{\bibfnamefont{N.}~\bibnamefont{Khaneja}},
  \bibinfo{author}{\bibfnamefont{B.}~\bibnamefont{Heitmann}},
  \bibinfo{author}{\bibfnamefont{A.}~\bibnamefont{Sp{\"o}rl}},
  \bibinfo{author}{\bibfnamefont{H.}~\bibnamefont{Yuan}},
  \bibinfo{author}{\bibfnamefont{T.}~\bibnamefont{Schulte-Herbr{\"u}ggen}},
  \bibnamefont{and} \bibinfo{author}{\bibfnamefont{S.~J.}
  \bibnamefont{Glaser}}, \bibinfo{journal}{Phys. Rev. A}
  \textbf{\bibinfo{volume}{75}}, \bibinfo{pages}{012322}
  (\bibinfo{year}{2007}).

\bibitem[{\citenamefont{Hall}(2010)}]{Hall2010}
\bibinfo{author}{\bibfnamefont{J.~A.~J.} \bibnamefont{Hall}},
  \bibinfo{journal}{Computational Management Science}
  \textbf{\bibinfo{volume}{7}}, \bibinfo{pages}{139} (\bibinfo{year}{2010}).

\bibitem[{\citenamefont{Jones}(2003)}]{Jones2003b}
\bibinfo{author}{\bibfnamefont{J.~A.} \bibnamefont{Jones}},
  \bibinfo{journal}{Phys. Rev. A} \textbf{\bibinfo{volume}{67}},
  \bibinfo{pages}{012317} (\bibinfo{year}{2003}).

\bibitem[{\citenamefont{Nielsen and Chuang}(2000)}]{NCbook}
\bibinfo{author}{\bibfnamefont{M.~A.} \bibnamefont{Nielsen}} \bibnamefont{and}
  \bibinfo{author}{\bibfnamefont{I.~L.} \bibnamefont{Chuang}},
  \emph{\bibinfo{title}{Quantum Computation and Quantum Information}}
  (\bibinfo{publisher}{CUP}, \bibinfo{year}{2000}).

\bibitem[{\citenamefont{Chuang and Nielsen}(1997)}]{Chuang1997}
\bibinfo{author}{\bibfnamefont{I.~L.} \bibnamefont{Chuang}} \bibnamefont{and}
  \bibinfo{author}{\bibfnamefont{M.~A.} \bibnamefont{Nielsen}},
  \bibinfo{journal}{J. Mod. Opt.} \textbf{\bibinfo{volume}{44}},
  \bibinfo{pages}{2455 } (\bibinfo{year}{1997}).

\bibitem[{\citenamefont{Bowdrey et~al.}(2002)\citenamefont{Bowdrey, Oi, Short,
  Banaszek, and Jones}}]{Bowdrey2002}
\bibinfo{author}{\bibfnamefont{M.~D.} \bibnamefont{Bowdrey}},
  \bibinfo{author}{\bibfnamefont{D.~K.~L.} \bibnamefont{Oi}},
  \bibinfo{author}{\bibfnamefont{A.~J.} \bibnamefont{Short}},
  \bibinfo{author}{\bibfnamefont{K.}~\bibnamefont{Banaszek}}, \bibnamefont{and}
  \bibinfo{author}{\bibfnamefont{J.~A.} \bibnamefont{Jones}},
  \bibinfo{journal}{Phys. Lett. A} \textbf{\bibinfo{volume}{294}},
  \bibinfo{pages}{258 } (\bibinfo{year}{2002}).

\bibitem[{\citenamefont{Nielsen}(2002)}]{Nielsen2002}
\bibinfo{author}{\bibfnamefont{M.~A.} \bibnamefont{Nielsen}},
  \bibinfo{journal}{Phys. Lett. A} \textbf{\bibinfo{volume}{303}},
  \bibinfo{pages}{249 } (\bibinfo{year}{2002}).

\bibitem[{\citenamefont{Hofmann}(2005)}]{Hofmann2005}
\bibinfo{author}{\bibfnamefont{H.~F.} \bibnamefont{Hofmann}},
  \bibinfo{journal}{Phys. Rev. Lett.} \textbf{\bibinfo{volume}{94}},
  \bibinfo{pages}{160504} (\bibinfo{year}{2005}).

\bibitem[{\citenamefont{Pedersen et~al.}(2007)\citenamefont{Pedersen,
  M{\o}ller, and M{\o}lmer}}]{Pedersen2007}
\bibinfo{author}{\bibfnamefont{L.~H.} \bibnamefont{Pedersen}},
  \bibinfo{author}{\bibfnamefont{N.~M.} \bibnamefont{M{\o}ller}},
  \bibnamefont{and}
  \bibinfo{author}{\bibfnamefont{K.}~\bibnamefont{M{\o}lmer}},
  \bibinfo{journal}{Phys. Lett. A} \textbf{\bibinfo{volume}{367}},
  \bibinfo{pages}{47 } (\bibinfo{year}{2007}).

\bibitem[{\citenamefont{Mayer and Knill}(2018)}]{Mayer2018}
\bibinfo{author}{\bibfnamefont{K.}~\bibnamefont{Mayer}} \bibnamefont{and}
  \bibinfo{author}{\bibfnamefont{E.}~\bibnamefont{Knill}},
  \bibinfo{journal}{Phys. Rev. A} \textbf{\bibinfo{volume}{98}},
  \bibinfo{pages}{052326} (\bibinfo{year}{2018}).

\bibitem[{\citenamefont{Kruschke}(2011)}]{Kruschke2011}
\bibinfo{author}{\bibfnamefont{J.~K.} \bibnamefont{Kruschke}},
  \emph{\bibinfo{title}{Doing Bayesian Data Analysis}}
  (\bibinfo{publisher}{Academic Press}, \bibinfo{year}{2011}).

\end{thebibliography}

\end{document}